\documentclass[prl,aps,superscriptaddress, twocolumn,floatfix,longbibliography]{revtex4-1}
\usepackage[utf8]{inputenc}

\usepackage{booktabs}
\usepackage{subcaption}
\usepackage{natbib}
\usepackage{graphicx}
\usepackage{color}
\usepackage[tbtags]{amsmath}
\usepackage[colorlinks, linkcolor=blue]{hyperref}
\usepackage{amssymb}
\usepackage{gensymb}
\usepackage{float}
\usepackage{amsmath}
\usepackage{tabularx,graphicx}
\usepackage{epstopdf}
\usepackage{latexsym}
\usepackage{color, colortbl}
\usepackage{psfrag}
\usepackage{bbm}
\usepackage{bm}
\usepackage{titlesec}
\usepackage{dsfont}
\usepackage{feynmp}
\usepackage{slashed}
\usepackage{multirow}
\usepackage{appendix}
\usepackage{ragged2e}
\usepackage{bbold}

\extrafloats{100}

\begin{document}

\title{Soft information from neural-network logits and correlation-aware matching gaps}

\author{David Dentelski}
\email{daviddentelski@gmail.com}

\date{\today}

\begin{abstract}
To utilize quantum error-correcting codes, a decoder must infer the logical sector from the measured syndrome. Beyond producing a hard logical decision, some decoders provide soft information that estimates the reliability of that decision. For minimum-weight perfect matching (MWPM), a common confidence measure is the complementary gap. Neural-network decoders trained with binary cross entropy naturally output a different confidence score, the logit, whose usefulness for post-selection has already been demonstrated.  Here, we compare the logit of a pretrained graph neural-network (GNN) decoder with the complementary gaps of uncorrelated and correlation-aware MWPM, evaluated on identical syndromes from rotated-surface-code memory experiments under circuit-level noise. Incorporating correlations substantially improves matching-based hard decisions, post-selection performance, and calibration, reversing the GNN's advantage at larger code distances while leaving it superior at smaller distance. Exchanging confidence scores shows that their ability to rank logical failures is decoder-relative. An exact-posterior comparison at small distance, together with larger-distance log-loss measurements, further shows that neither score universally provides the closer approximation to posterior soft information. These results reveal that much of the apparent advantage of learned confidence over the standard matching gap arises from correlations omitted by the matching model. Nevertheless, learned logits provide useful soft information without requiring an explicit detector error model or competing-sector matching calculations at inference, offering an alternative where matching-based confidence is costly or unavailable.
\end{abstract}

\maketitle

\section{Introduction}
Soft information refers to decoder output that provides not only a binary decision about the logical sector, but also a confidence measure for that decision. Such information can be passed to another decoding layer in concatenated codes \cite{poulin2006optimal, pattison2025hierarchical, haruna2025hierarchical}, or used for post-selection, where low-confidence runs are discarded to reduce the logical error rate \cite{knill2005quantum, aliferis2007accuracy}. Post-selection has long been used to improve non-fault-tolerant operations, particularly magic-state injection \cite{li2015magic, chamberland2020very, singh2022high, gidney2023cleaner}, distillation \cite{bravyi2005universal, bravyi2012magic, litinski2019magic, lee2025low}, and cultivation \cite{gidney2024magic, rosenfeld2025magic, sahay2025fold, claes2025cultivating, vaknin2026high, chen2026efficient}. More recently, related ideas have been extended to broader fault-tolerant protocols, where mid-circuit syndrome information is used to discard faulty runs or resource states \cite{prabhu2024distance, dobbs2025advantage, chen2025scalable, bhambay2026adaptive, akahoshi2025runtime, staples2026scalable}. Several variants of post-selection can be used, depending on which information is extracted from the measured syndromes \cite{bombin2024fault}.

For a binary logical observable, the ideal soft output is the posterior log-likelihood ratio (LLR) of the logical sector given the syndrome, Eq.~(\ref{true_gap}) below. Practical decoders approximate it in different ways. For minimum-weight perfect matching (MWPM), the standard confidence measure is the complementary gap \cite{bombin2024fault, meister2024efficient, gidney2025yoked, akahoshi2025runtime, dincua2025error}: the weight difference between the minimum-weight correction in the predicted logical sector and that in the complementary sector. Its empirical relation to the logical failure probability is well described by a logistic form with a fitted, setting-dependent slope \cite{gidney2025yoked, akahoshi2025runtime, staples2026scalable}, and it has been compared with other matching-based confidence measures, such as the swim distance \cite{dincua2025error}. As an estimate of the posterior LLR, however, the gap has two known limitations. It retains only the minimum-weight configuration in each sector, and thus ignores their multiplicity, and it is only as accurate as the error model encoded in the matching graph. In particular, correlations between X- and Z-type detection events created by Y-type errors are discarded by standard matching, unless they are reintroduced, for example by the two-pass reweighting of correlated matching \cite{fowler2013optimal, pymatchingv2}. Evaluating the gap also requires comparing competing logical sectors, whose number grows as $4^{k}$ for a code with $k$ logical qubits \cite{lee2026efficient}, which has motivated cheaper constructions \cite{wills2026forced, kishi2026even} and gap-like quantities for partially available syndromes \cite{staples2026scalable}.

Neural-network (NN) and related learning-based decoders obtain soft information through a different route. They learn a map from syndromes to logical labels directly from simulated or experimental data \cite{torlai2017neural, krastanov2017deep, baireuther2019neural, meinerz2022scalable, bausch2024learning, zhong2024advantage, lange2025data, senior2025scalable, gu2026scalable}, and in many architectures the decision is the sign of a scalar logit. When trained with binary cross entropy (BCE), this logit is driven toward the posterior log-odds of the label under the training distribution \cite{seip2026machine}, and learned outputs have been used as calibrated confidence estimates and for post-selection \cite{bausch2024learning, gu2026scalable}, including diffusion-model decoding \cite{xu2026diffqec} and learned decoder-independent syndrome scores \cite{haug2026machine}. The logit is produced in the same forward pass as the decision, and no explicit comparison of logical sectors is required at inference. The training objective alone, however, does not guarantee that its magnitude is a reliable confidence: the decision depends only on its sign, and a network can reduce its error rate while still assigning high confidence to some of its failures.

Comparing a learned logit with a matching gap therefore raises a question that goes beyond which of the two performs better: \textit{what determines the quality of decoder soft information?} At least four factors are entangled in such a comparison. (i) Hard-decision quality: a confidence score is evaluated against the failures of the decision it accompanies, so decoders with different failure sets face different ranking problems. (ii) The error model available to the decoder: a matching graph without correlations discards information that a network, which sees both X- and Z-type detection events, can exploit. (iii) Minimum-weight versus posterior information: even with the correct error model, a matching gap ignores the multiplicity of configurations within each sector. (iv) The learned model itself: a network approximates the posterior of the distribution it was trained on, with finite capacity and finite training. 

Here we separate these factors for the pretrained graph neural-network (GNN) decoder of Ref.~\cite{lange2025data} on the rotated surface code under uniform circuit-level noise. We evaluate three decoders on identical sampled syndromes, with $10^8$ shots per configuration: the GNN, standard uncorrelated MWPM, and correlation-aware MWPM, whose complementary gap is computed on a shared, shot-conditioned reweighted matching graph \cite{gidney2025yoked}. Switching the correlations on and off within the same matching framework isolates factor (ii). Exchanging the confidence scores between the decoders probes how the ranking depends on the hard decision, factor (i). Calibration probes the scale of each score. Finally, at distance $d = 3$ we compute the exact posterior of every sampled syndrome, together with the posterior of the GNN's training distribution. This provides a ground-truth posterior against which the combined effects of factors (iii) and (iv) can be studied, although it does not isolate them. At larger distances, differences in the native log-loss estimate differences in the mean divergence from the same posterior, without requiring the posterior to be computed.

We find that correlations account for a large part of the difference between the two kinds of soft output. Including them improves the hard decisions of matching and, at the same time, the ranking and the calibration of its gap. With correlations, the native matching decoder-gap pair overtakes the GNN decoder-logit pair at $d = 7$ and $9$, by a margin that grows with the distance and with the strength of post-selection, whereas the GNN-logit pair remains better at $d = 5$. Exchanging the scores shows that discrimination is decoder-relative: each score ranks the failures of its own decoder best, apart from a small residual reversal at the smallest configuration. The calibration disadvantage of the standard matching gap largely disappears once correlations are included, and at $d \geq 7$ the native log-loss, whose expectation equals the mean divergence from the posterior up to a constant common to all decoders, shows that the correlated gap is on average closer to the posterior than the logit. At $d = 3$, the GNN logit is remarkably close to the exact posterior, and a substantial part of its remaining deviation is consistent with the mismatch between the posterior of its training mixture and the fixed-$p$ posterior, whereas the correlated gap retains a residual concentrated near the decision boundary. Together, these results indicate that the quality of decoder soft information is determined to a large extent by the statistical information that the decoder model represents.

\section{Decoders and confidence scores}
\phantomsection\label{sec:model}

\emph{Posterior target.} Let $s$ denote the measured syndrome and $L \in \{0, 1\}$ the logical sector. The posterior LLR is
\begin{align}\label{true_gap}
    \Lambda(s) = \ln \dfrac{P(L = 0 | s)}{P(L = 1| s)}.
\end{align}
Its sign gives the most likely sector, the maximum a posteriori (MAP) decision, and its magnitude $|\Lambda(s)|$ is the ideal confidence in that decision. We compare two practical confidence scores, defined below: the magnitude of the GNN logit and the magnitude of the MWPM complementary gap. For the signed-confidence distributions, we also define $\Delta_D$ as the confidence of decoder $D$, positive when its decision is correct and negative otherwise. This sign requires the true logical outcome, which is known in simulation but not during decoding. Confidences are quoted in dB, with $1$ nat $= 10/\ln 10 \approx 4.34$ dB. To avoid ambiguity, we reserve ``gap'' for the matching quantities defined below and ``posterior LLR'' for $\Lambda$.

\emph{GNN decoder and its training distribution.} The architecture and training procedure of the GNN are described in Ref.~\cite{lange2025data}; we review only the elements relevant here. The input is a graph whose nodes are the detectors that fired in a shot, each annotated by a five-dimensional feature vector: three space-time coordinates and two binary indicators of whether the detector belongs to an X- or a Z-type stabilizer. Edges connect nearby nodes, with weights determined by their Euclidean distance, and are pruned to a fixed maximum node degree, so that the graph encodes the spatio-temporal structure of the syndrome without using detector error model (DEM) weights. Graph-convolution layers aggregate information from neighboring detectors, a global pooling maps the graph to a fixed-size embedding, and a feed-forward network outputs a single logit $z(s)$. We orient $z$ such that $z > 0$ corresponds to the decision $\hat{L} = 0$, as for $\Lambda$, and use $g_{\rm GNN} = |z(s)|$ as the confidence. We use the released checkpoints, one network per configuration $(d, r)$, where $d$ is the code distance and $r$ the number of syndrome-extraction rounds; all networks share the same architecture, except for a larger variant at $d = 9$. They were trained with BCE on an even mixture of syndromes generated at $p \in \{1, 2, 3, 4, 5\} \times 10^{-3}$, with freshly generated syndromes for every batch ($10^7$ samples per epoch) and with trivial syndromes excluded; the physical error rate $p$ is not an input to the network \cite{lange2025data}. Since BCE is a proper scoring rule, its population minimizer is the posterior log-odds of the training distribution \cite{seip2026machine}, which for this mixture is
\begin{align}\label{mixture}
    \Lambda_{\rm mix}(s) = \ln \dfrac{\sum_{p} w_p\, P(s, L = 0 | p)}{\sum_{p} w_p\, P(s, L = 1 | p)},
    \qquad w_p = \dfrac{1}{1 - P(\emptyset | p)},
\end{align}
where $P(\emptyset | p)$ is the probability of the trivial syndrome. Even an ideally trained network therefore targets $\Lambda_{\rm mix}$, and not the fixed-$p$ posterior of a particular benchmark. At the upper edge of the training range, $p = 5 \times 10^{-3}$, where most of our benchmarks are taken, the other components of the mixture are less noisy and typically assign more confident posteriors to the same syndrome, so that $\Lambda_{\rm mix}$ is expected to overstate the confidence. We quantify this effect exactly at $d = 3$ below (Fig.~\ref{fig:exact_d3}). Trivial syndromes, for which no detector fires, bypass the network at inference: they are assigned the decision $\hat{L} = 0$ and, as confidence, the MWPM gap of the trivial syndrome, so that both decoders carry the same confidence on these shots (\hyperref[app:robustness]{Appendix~E}).

\emph{Matching gaps.} For MWPM, the DEM generated by \texttt{Stim} \cite{gidney2021stim}, decomposed into graph-like components, defines a matching graph with edge weights $\omega_e = \ln[(1 - p_e)/p_e]$, where $p_e$ is the probability of the error mechanism on edge $e$. The signed complementary gap is \cite{bombin2024fault}
\begin{align}\label{MWPM_gap}
    g = \omega(l_{\rm wrong}) - \omega(l_{\rm correct}),
\end{align}
where $\omega(l) = \sum_{e \in l} \omega_e$ is the weight of a correction $l$, and $l_{\rm correct}$ and $l_{\rm wrong}$ are the minimum-weight corrections in the correct and in the incorrect logical sectors, respectively. We compute both by promoting the logical observable to a virtual detector and matching twice, with this detector forced to $0$ and to $1$ \cite{gidney2025yoked}. With this convention, $g = \Delta_{\rm MWPM}$ is positive when the decoder, which selects the sector of smaller minimum weight, succeeds; the confidence accessible in an experiment is its magnitude $|g|$. We consider two matching decoders, both based on \texttt{PyMatching} \cite{higgott2021pymatching, pymatchingv2}. For the uncorrelated decoder, MWPM$_{\rm uncorr}$, the components of each error mechanism are treated as independent edges, and the resulting gap is $g_{\rm uncorr}$. The correlation-aware decoder, MWPM$_{\rm corr}$, uses the two-pass reweighting of correlated matching \cite{fowler2013optimal, pymatchingv2}. A first, uncorrelated matching pass identifies a set of candidate error edges $E$. For each $e \in E$, correlated partner edges $a$ arising from the same decomposed error mechanism are reweighted using the conditional probability $q = \min\{1/2, P(e \wedge a)/P(e)\}$, replacing their weight by $\ln[(1-q)/q]$ whenever this is smaller. A second matching pass on the resulting shot-dependent graph gives the correlated decoding decision. Following the complementary-gap construction of Ref.~\cite{gidney2025yoked}, we compute $g_{\rm corr}$ by forcing the virtual logical detector to each of its two values and comparing the minimum matching weights on this same reweighted graph. The graph is constructed from a single first pass, unconstrained with respect to the logical sector, rather than reweighted separately for each logical sector. The sector of smaller minimum weight reproduces the correlated decision of \texttt{PyMatching}, as verified on all $10^8$ shots of every configuration, apart from exact ties (\hyperref[app:corr]{Appendix~B}). This construction matters: reweighting each logical sector separately from a sector-constrained first pass, as in Ref.~\cite{pymatchingrs}, compares minimum weights on different graphs and yields substantially worse post-selection performance in our tests (\hyperref[app:corr]{Appendix~B}). In the remainder of the paper, ``MWPM'' denotes MWPM$_{\rm corr}$ unless it is explicitly labeled as uncorrelated; MWPM$_{\rm uncorr}$ and $g_{\rm uncorr}$ serve as the ablation without correlations.

\emph{Simulations and evaluation.} We simulate Z-basis memory experiments of the rotated surface code with \texttt{Stim} \cite{gidney2021stim}, under the uniform circuit-level depolarizing noise used to train the GNN \cite{lange2025data}, with strength $p$ after every Clifford gate, on the data qubits before every round, and on every measurement and reset. At $p = 5 \times 10^{-3}$ we simulate the eight configurations $(d, r) \in \{(5, 5), (5, 7), (5, 9), (5, 11), (7, 7), (7, 9), (7, 11), (9, 9)\}$ with $10^8$ shots each, and at $d = r \in \{5, 7, 9\}$ also $p \in \{3, 3.5, 4, 4.5\} \times 10^{-3}$ with $10^8$ shots each.
All decoders are evaluated on the identical shots. For post-selection, the shots are ordered by a confidence score and the fraction $\kappa$ with the highest confidence is accepted; $L_D(\kappa)$ denotes the post-selected logical error rate (LER) of decoder $D$, i.e., the fraction of accepted shots on which $D$ fails. Unless stated otherwise, each decoder is ranked by its own confidence, which we call the native pair. Statistical conventions, tie handling, and the paired resampling used for all ratios are described in \hyperref[app:stats]{Appendix~A}. The code that generates the shot data, the correlated gap, all figures and tables, and the numbers quoted in the text is publicly available \cite{dentelski2026code}. 

\begin{figure*}[t]
    \centering
    \includegraphics[width=0.85\textwidth]{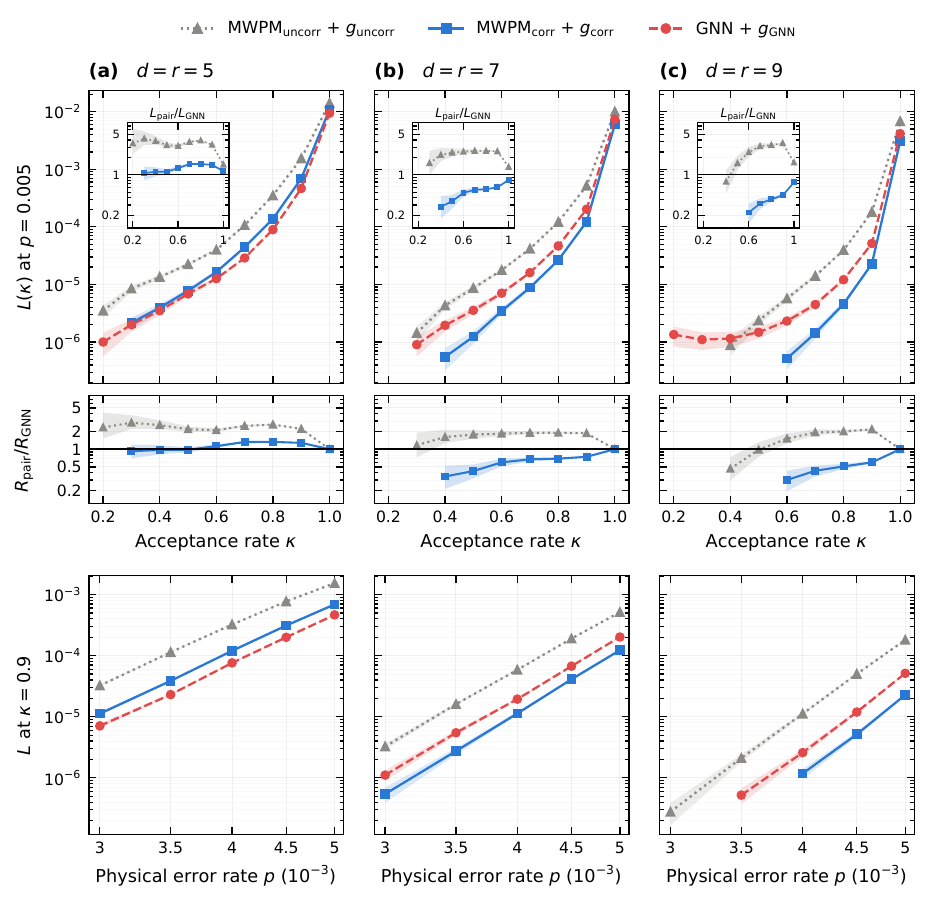}
    \caption{\textbf{Post-selection with the native confidence of each decoder.} Rotated surface code, Z-memory, with (a) $d = r = 5$, (b) $d = r = 7$, and (c) $d = r = 9$. Each decoder is ranked by its own confidence score: MWPM$_{\rm uncorr}$ by $g_{\rm uncorr}$ (gray triangles, dotted), MWPM$_{\rm corr}$ by $g_{\rm corr}$ (blue squares, solid), and the GNN by its logit (red circles, dashed). Top: post-selected LER $L(\kappa)$ as a function of the acceptance rate $\kappa$ at $p = 5 \times 10^{-3}$; the insets retain the overall ratio $L_{\rm pair}/L_{\rm GNN}$ of each matching pair to the GNN pair, with paired $95\%$ intervals (\hyperref[app:stats]{Appendix~A}); values below $1$ favor the matching pair. Middle: ratio $R_{\rm pair}/R_{\rm GNN}$ of the selection gains $R_D(\kappa) = L_D(\kappa)/L_D(1)$, with paired $95\%$ intervals; values below $1$ indicate a larger relative improvement from post-selection for the matching pair. Bottom: $L$ at $\kappa = 0.9$ as a function of $p$. Each point uses $10^8$ shots, the same shots for all decoders. Shaded bands are $95\%$ confidence intervals. A point is shown only where it rests on at least $20$ accepted failures (for the ratios, in both pairs); the counts behind the top row, including the suppressed points, are listed in Table~\ref{counts_table}.}
\label{fig:LER}
\end{figure*}

\section{Results}

\subsection{Native decoder-confidence pairs and the correlation ablation}
\phantomsection\label{sec:native}

We start by comparing the decoders paired with their native confidence scores. Fig.~\ref{fig:LER} (top) 
shows the post-selected LER as a function of the acceptance rate $\kappa$, for $d = r = 5, 7, 9$. Compared with standard uncorrelated MWPM, the GNN achieves a lower LER at every distance for $\kappa \gtrsim 0.5$. However, incorporating correlations into MWPM changes this picture: the correlated matching-gap pair outperforms the GNN-logit pair at $d = 7$ and $9$, while the GNN retains its advantage at $d = 5$. At $d = 7$ and $9$, these differences become more pronounced as post-selection becomes stronger.  Thus, the apparent superiority of learned confidence depends substantially on the information represented by the matching model.

To distinguish the contributions of hard-decision performance and post-selection, we consider the ratio of the native decoder-confidence pairs, shown in the insets of Fig.~\ref{fig:LER}. Defining the selection gain as $R_D(\kappa) = L_D(\kappa)/L_D(1)$ for decoder $D$, we write
\begin{align}\label{eq:decomposition}
\frac{L_{\rm MWPM}(\kappa)}{L_{\rm GNN}(\kappa)}
= \frac{L_{\rm MWPM}(1)}{L_{\rm GNN}(1)}
\times \frac{R_{\rm MWPM}(\kappa)}{R_{\rm GNN}(\kappa)}.
\end{align}
The first factor compares the hard-decision error rates, while the second, shown in the middle panels of Fig.~\ref{fig:LER}, measures the relative improvement due to post-selection.  At $\kappa = 0.9$ the decomposition gives $1.48 = 1.16 \times 1.28$ at $d = 5$, $0.61 = 0.82 \times 0.74$ at $d = 7$, and $0.44 = 0.74 \times 0.60$ at $d = 9$. The two factors point in the same direction at every distance: where correlated matching makes fewer errors, its gap also extracts a larger selection gain from them, and the selection factor departs further from unity than the hard-decision factor.

The uncorrelated pair (gray) highlights the role of correlations. At $d = 9$ and $\kappa = 0.9$, incorporating correlations reduces the post-selected LER by a factor of $7.9$ relative to uncorrelated MWPM. Thus, much of the apparent advantage of the GNN over standard MWPM arises from correlations between $X$- and $Z$-type detection events that the uncorrelated matching graph neglects, rather than necessarily from an intrinsically better confidence measure.

To test the robustness of this comparison, Fig.~\ref{fig:LER} (bottom) shows the post-selected LER at $\kappa=0.9$ for $p=3$--$5\times10^{-3}$. Interestingly, as $p$ decreases, the hard-decision advantage of correlated MWPM over the GNN narrows at larger distances, while its post-selection advantage persists. This further highlights the contribution of confidence-based selection beyond hard-decision performance. Recent work has shown that correlated matching can lose effective distance at sufficiently low physical error rates \cite{hernando2026correlated}. We see no indication of this within the tested range, where correlated MWPM improves relative to uncorrelated matching as $p$ decreases. Nevertheless, the effect reported in Ref.~\cite{hernando2026correlated} occurs at lower error rates than studied here, and our observations do not establish the asymptotic behavior. Whether the hard-decision ordering persists further below threshold remains open.

\subsection{Exchanged scores: discrimination is decoder-relative}
\phantomsection\label{sec:crossed}

\begin{figure*}
    \centering
    \includegraphics[width=0.95\textwidth]{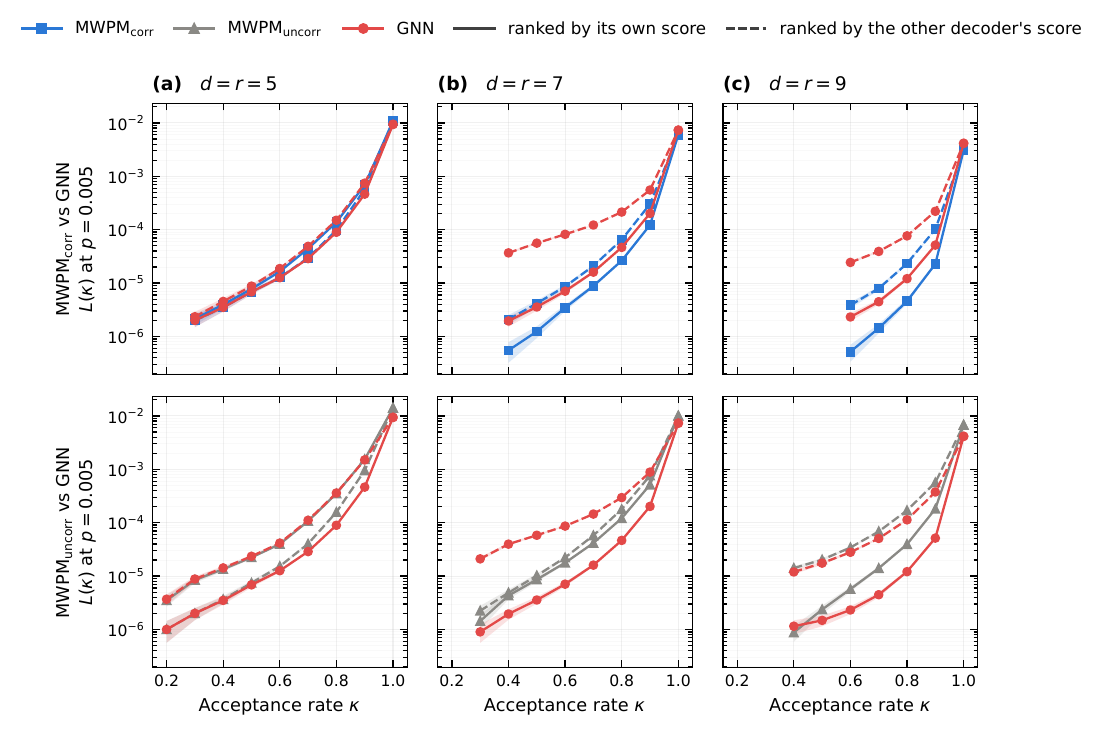}
    \caption{\textbf{Post-selection with exchanged confidence scores.} Post-selected LER as a function of the acceptance rate $\kappa$ at $p = 5 \times 10^{-3}$, for (a) $d = r = 5$, (b) $d = r = 7$, and (c) $d = r = 9$, computed from the same $10^8$ shots per configuration. Top row: MWPM$_{\rm corr}$ (blue squares) and the GNN (red circles); bottom row: the same exchange with MWPM$_{\rm uncorr}$ (gray triangles). Solid curves show each decoder ranked by its own confidence score, dashed curves the same decoder ranked by the magnitude of the other decoder's score. Curves of equal color share the identical set of logical failures and differ only in the order in which shots are discarded; they coincide at $\kappa = 1$. Shaded bands are $95\%$ confidence intervals. A $\kappa$ point is shown only where all four curves of the panel rest on at least $20$ accepted failures (Tables~\ref{counts_table} and \ref{crossed_counts_table}).}
\label{fig:crossed}
\end{figure*}

To examine the role of confidence ranking separately from hard-decision performance, we exchange the confidence scores while keeping each decoder's decisions fixed. Since both scores are evaluated on the same syndromes, each decoder retains exactly the same successes and failures, but the order in which shots are discarded changes. This allows us to compare how effectively each score identifies the failures of a given decoder.

Fig.~\ref{fig:crossed} (top) reveals a clear pattern: at $d=7$ and $9$, both exchanged (dashed) curves lie above their native partners at every displayed acceptance rate. That is to say that exchanging the confidence scores consistently worsens post-selection for both decoders. Although both scores are computed from the same syndromes, each identifies the failures of its own decoder more effectively. At $d=5$, however, the GNN logit slightly outperforms the correlated gap even when ranking MWPM's failures, suggesting that this behavior is not universal.

An important subtlety arises when the decoders disagree. A large confidence score then expresses confidence in the opposite logical sector, yet exchanging only its magnitude treats the shot as reliable for the other decoder. Such disagreements occur on at most $1.7\%$ of shots (below $1\%$ at $d = r$), but are concentrated among failures (\hyperref[app:pairing]{Appendix~D}). To assess their role, we also consider a decision-aware exchange, assigning the transferred confidence a positive sign when the decisions agree and a negative sign when they differ. This joint two-decoder statistic preserves the sector information that is lost when only confidence magnitudes are exchanged.

To quantify the ranking independently of a particular acceptance rate, Table~\ref{auc_table} reports the area under the receiver-operating-characteristic curve (AUC): the probability that a randomly selected correct decision receives a higher confidence than a failure, with ties counted as half. At $d \geq 7$, each native score achieves the higher AUC for its own decoder, while at $d = 5$ the only resolved reversal occurs at $(5,5)$. Remarkably, accounting for disagreements recovers most of the exchanged-ranking penalty at larger distances, and the decision-aware correlated gap even surpasses the native GNN logit when ranking GNN failures at $d \geq 7$. Thus, the poor performance of exchanged confidence magnitudes partly reflects the loss of information about the predicted logical sector, rather than simply an inferior confidence ranking.

The comparison with uncorrelated matching, shown in the bottom panels of Fig.~\ref{fig:crossed}, reinforces the role of the error model. Without correlations, the GNN logit ranks MWPM's failures better throughout the $d = 5$ configurations, while transferring the matching gap to the GNN causes a substantially larger penalty than with the correlated gap. Incorporating correlations improves the gap's ability to identify not only MWPM's own failures but also those of the GNN. Interestingly, correlations also increase the overlap between the decoders' failures. At $d = 9$, for example, the fraction of matching failures shared with the GNN rises from $28\%$ to $46\%$. Nevertheless, the two scores still produce distinct rankings, motivating a closer examination of where their successes and failures lie along their respective confidence scales.

\begin{figure*}
    \centering
    \includegraphics[width=0.9\textwidth]{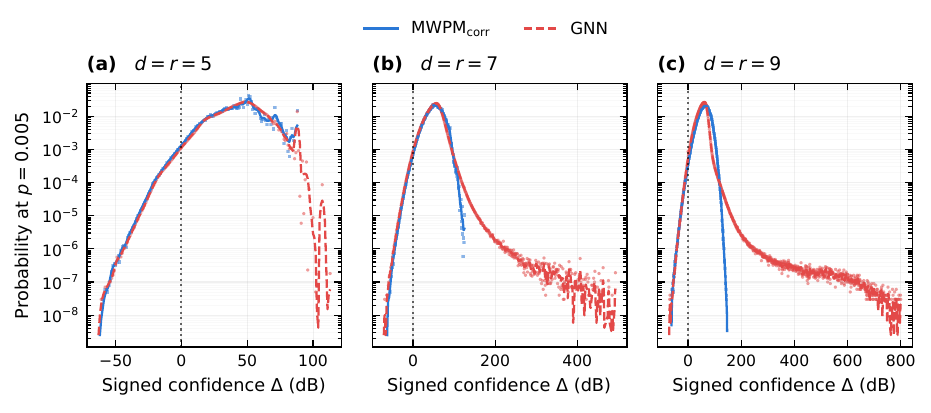}
    \caption{\textbf{Distributions of the signed confidence} $\Delta_D$ of MWPM$_{\rm corr}$ (blue, solid) and of the GNN (red, dashed) at $p = 5 \times 10^{-3}$, $10^8$ shots, for (a) $d = r = 5$, (b) $d = r = 7$, and (c) $d = r = 9$. Positive values correspond to correctly decoded shots and negative values to logical failures. Points are integer-dB bins with at least three shots; the lines are smoothed by convolution with a cosine window. The gap cannot exceed the weighted distance of the code, $88.2$, $125.5$ and $162.7$ dB (the gap of the trivial syndrome); the GNN logit is not bounded in this way. The confidence axis is truncated for clarity.}
\label{fig:gap1}
\end{figure*}

Fig.~\ref{fig:gap1} provides another perspective on the exchanged-score results, showing where correctly decoded shots and failures lie along each decoder's signed-confidence scale. Both scores concentrate failures at low confidence, with similar distributions around the peak and on the failure side. The main difference appears in the high-confidence tail: the matching gap is bounded by the weighted distance of the code, while the GNN logit extends to much larger values. This raises the possibility that saturation of the gap explains the reversal observed at $(5,5)$. Indeed, for the uncorrelated gap, $6.7\%$ of shots accumulate at its upper bound. However, incorporating correlations reduces this fraction to $1.5\%$, almost entirely trivial syndromes, while the reversal persists. The gap ceiling is therefore not responsible for this exception.

Taken together, these results show that confidence discrimination is decoder-relative. Although both scores are derived from the same syndrome, neither provides a universally better ranking of decoder failures. At larger distances, each score best identifies the failures of its own decoder, while the remaining exception at $(5,5)$ is small once correlations are included. Moreover, the decision-aware exchange reveals that decoder disagreements contain valuable additional information: signing the correlated gap by agreement ranks the GNN's failures better than its native logit at $d\geq7$ (Table~\ref{auc_table}). \hyperref[app:pairing]{Appendix~D} examines these differences in terms of shared and unshared failures.

\begin{table*}
\centering
\begin{tabular}{ll|cccccccc}
\toprule
Decoder & Ranked by & $(5, 5)$ & $(5, 7)$ & $(5, 9)$ & $(5, 11)$ & $(7, 7)$ & $(7, 9)$ & $(7, 11)$ & $(9, 9)$ \\
\hline
\midrule
MWPM$_{\rm corr}$ & $g_{\rm corr}$ & 0.9753 & \textbf{0.9680} & \textbf{0.9610} & \textbf{0.9545} & \textbf{0.9861} & \textbf{0.9826} & \textbf{0.9793} & \textbf{0.9922} \\
MWPM$_{\rm corr}$ & $|g_{\rm GNN}|$ & \textbf{0.9767} & 0.9679 & 0.9600 & 0.9519 & 0.9743 & 0.9663 & 0.9582 & 0.9799 \\
MWPM$_{\rm corr}$ & $\pm g_{\rm GNN}$ (signed) & 0.9826 & 0.9774 & 0.9725 & 0.9678 & 0.9844 & 0.9803 & 0.9753 & 0.9904 \\
\hline
GNN & $g_{\rm GNN}$ & \textbf{0.9785} & \textbf{0.9719} & \textbf{0.9658} & \textbf{0.9598} & \textbf{0.9838} & \textbf{0.9794} & \textbf{0.9749} & \textbf{0.9898} \\
GNN & $|g_{\rm corr}|$ & 0.9692 & 0.9592 & 0.9499 & 0.9407 & 0.9691 & 0.9636 & 0.9513 & 0.9763 \\
GNN & $\pm g_{\rm corr}$ (signed) & 0.9733 & 0.9665 & 0.9598 & 0.9537 & 0.9904 & 0.9883 & 0.9871 & 0.9953 \\
\hline
MWPM$_{\rm uncorr}$ & $g_{\rm uncorr}$ & 0.9668 & 0.9557 & 0.9452 & 0.9353 & \textbf{0.9775} & \textbf{0.9711} & \textbf{0.9650} & \textbf{0.9842} \\
MWPM$_{\rm uncorr}$ & $|g_{\rm GNN}|$ & \textbf{0.9747} & \textbf{0.9642} & \textbf{0.9547} & \textbf{0.9447} & 0.9706 & 0.9609 & 0.9505 & 0.9693 \\
\bottomrule
\end{tabular}
\caption{Tie-aware AUC with which each score ranks the logical failures of each decoder, at $p = 5 \times 10^{-3}$ ($10^8$ shots per configuration $(d, r)$, the same shots for all decoders; ties receive half credit). For each decoder, the rows give its own score, the magnitude of the other decoder's score (bold: the better of these two), and the other decoder's score signed by agreement ($+|g|$ where the two decisions agree, $-|g|$ where they differ), which is a two-decoder statistic. The last two rows give the failures of MWPM$_{\rm uncorr}$, ranked by its own gap and by the logit. The paired standard errors of same-decoder AUC differences are below $10^{-4}$ (\hyperref[app:robustness]{Appendix~E}).}
\label{auc_table}
\end{table*}

\subsection{Ground truth: the exact posterior at \texorpdfstring{$d = 3$}{d = 3}}
\phantomsection\label{sec:exact}

\begin{figure*}
    \centering
    \includegraphics[width=0.9\textwidth]{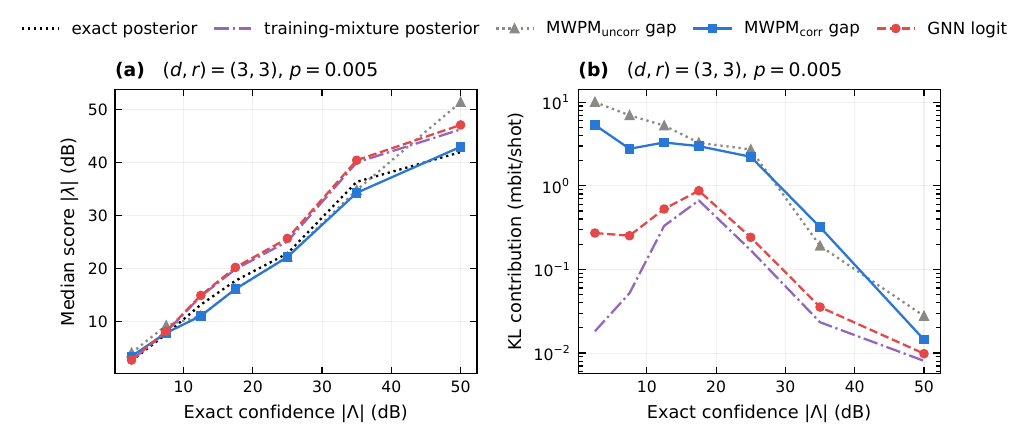}
    \caption{\textbf{Confidence scores compared with the exact posterior} at $(d, r) = (3, 3)$, $p = 5 \times 10^{-3}$, on $57{,}444{,}877$ non-trivial shots, binned by the exact confidence $|\Lambda(s)|$ (bins $0$--$5$, $5$--$10$, $10$--$15$, $15$--$20$, $20$--$30$, $30$--$40$, and $40$--$60$ dB, plotted at their centers). (a) Median magnitude of each score in each bin; the black dotted line is the exact posterior itself and the dash-dotted line the posterior $\Lambda_{\rm mix}$ of the GNN's training mixture, Eq.~(\ref{mixture}). (b) Contribution of each bin to the mean Kullback-Leibler divergence of $q(L = 0|s) = \sigma(\lambda(s))$ from the exact posterior, with $\lambda$ the score signed toward its decision (\hyperref[app:exact]{Appendix~F}), in mbit per shot (the contributions add up to the totals in Table~\ref{exact_d3_table}).}
\label{fig:exact_d3}
\end{figure*}

\begin{table}
\centering
\footnotesize
\setlength{\tabcolsep}{2pt}
\begin{tabular}{l|c|cc|c}
\toprule
Score & LER & KL ($a = 1$) & KL (best $a$) & KL to $\Lambda_{\rm mix}$ \\
\hline
\midrule
exact $\Lambda$ & 2.513e-02 & 0.00$\pm$0.00 & 0.00 (1.000) & 1.08 \\
$\Lambda_{\rm mix}$ & 2.514e-02 & 1.27$\pm$0.00 & 0.19 (0.902) & 0.00 \\
MWPM$_{\rm uncorr}$ & 2.989e-02 & 28.49$\pm$0.02 & 28.15 (0.948) & 30.92 \\
MWPM$_{\rm corr}$ & 2.895e-02 & 16.97$\pm$0.01 & 16.80 (1.040) & 19.84 \\
GNN & 2.530e-02 & 2.22$\pm$0.00 & 0.79 (0.889) & 0.66 \\
\bottomrule
\end{tabular}
\caption{Exact-posterior benchmark at $(d, r) = (3, 3)$, $p = 5 \times 10^{-3}$ ($57{,}444{,}877$ non-trivial shots): LER of the decision ${\rm sign}(\lambda)$, and mean per-shot Kullback-Leibler divergence (mbit per shot) of $q(L = 0|s) = \sigma(a \lambda(s))$ from the exact posterior, natively and with the best global rescaling $a$ (in parentheses), and from the posterior $\Lambda_{\rm mix}$ of the GNN's training mixture, Eq.~(\ref{mixture}).}
\label{exact_d3_table}
\end{table}

The exchanged-score analysis reveals which score better identifies each decoder's failures, but does not tell us how closely either score approximates the posterior. To address this question directly, we turn to the smallest code, where the posterior can be computed exactly. At $(d,r) =(3,3)$, the $24$ detectors allow us to obtain $P(s, L)$ for all $2^{25}$ syndrome-observable pairs from the full DEM ($219$ error mechanisms) using a Walsh-Hadamard transform (\hyperref[app:exact]{Appendix~F}). We can therefore compare the signed confidence scores with both the exact posterior $\Lambda(s)$ and the training-mixture posterior $\Lambda_{\rm mix}(s)$ of Eq.~(\ref{mixture}). We quantify the discrepancy using the mean per-shot Kullback-Leibler (KL) divergence, both at the native confidence scale and after optimizing a single global rescaling.

The results in Table~\ref{exact_d3_table} reveal a striking difference. At their native scales, the uncorrelated and correlated gaps have divergences of $28.5$ and $17.0$ mbit per shot, respectively, compared with only $2.2$ mbit for the GNN logit. Incorporating correlations thus removes about $40\%$ of the gap's divergence, but leaves a substantial residual. Importantly, this residual cannot be explained by a simple error in scale: optimizing a global rescaling reduces it by only $1\%$. Fig.~\ref{fig:exact_d3} shows that the discrepancy is especially pronounced near the decision boundary, where several competing error configurations may contribute to the posterior. Their multiplicity is absent from a minimum-weight gap, providing one possible explanation. However, the correlated reweighting is itself approximate, so we cannot isolate multiplicity as the sole cause.

The GNN, by contrast, comes remarkably close to the exact posterior. Its hard-decision LER nearly matches that of the MAP decoder ($2.530\%$ versus $2.513\%$), and a global rescaling reduces its divergence to just $0.79$ mbit per shot. Why, then, does its native logit still deviate from the fixed-$p$ posterior? The training mixture offers a natural explanation. The logit is only $0.66$ mbit away from $\Lambda_{\rm mix}$, and its confidence dependence closely follows that posterior rather than $\Lambda(s)$ (Fig.~\ref{fig:exact_d3}(a)). The mixture itself differs from the fixed-$p$ posterior, producing an overconfidence consistent in direction and approximate magnitude with that of the GNN. Thus, at $d = 3$, the network appears to have learned a close approximation to the posterior it was trained to predict. The agreement does not establish an exact decomposition of the remaining discrepancy, since KL divergences are not additive.

Yet this near-optimal behavior at $d = 3$ does not extend to larger codes. At $d \geq 7$, correlated matching outperforms the GNN both in failure ranking (Fig.~\ref{fig:crossed} and Table~\ref{auc_table}) and in estimated mean divergence from the posterior, as inferred from the log-loss identity in Eq.~(\ref{eq:logloss}) below. This raises an important question: does the network approximate its training-mixture posterior less accurately as the code grows, or does the correlated gap become a better approximation to the true posterior? A changing influence of the training mixture may also contribute, but we have not directly tested this possibility. Since exact posterior calculations are impractical at these larger distances, we cannot distinguish these explanations. The contrast nevertheless demonstrates that excellent soft-output performance at small distance does not necessarily predict performance at larger codes.

\subsection{Calibration}
\phantomsection\label{sec:calibration}

\begin{figure*}
    \centering
    \includegraphics[width=0.9\textwidth]{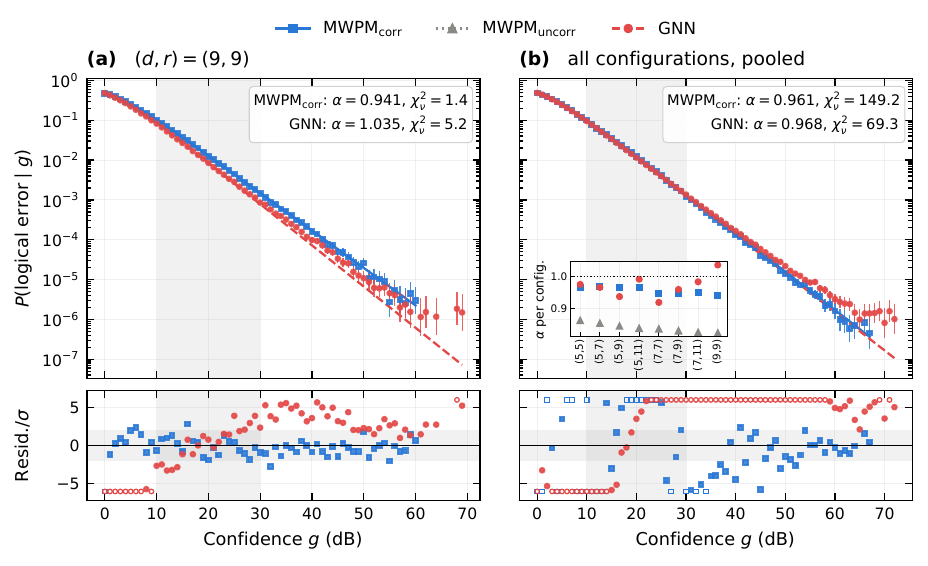}
    \caption{\textbf{Empirical calibration of the confidence scores} at $p = 5 \times 10^{-3}$. (a) Conditional logical-failure probability of MWPM$_{\rm corr}$ and of the GNN as a function of their confidence $g$ at $(d, r) = (9, 9)$ (points, with $95\%$ Wilson intervals), with maximum-likelihood fits of Eq.~(\ref{true_gap_dB}) (lines). The fits maximize the binomial likelihood of counts aggregated on a $0.001$-dB grid over the window $g \in [10, 30]$ dB (shaded), which was fixed in advance and is common to all configurations and decoders (other windows: \hyperref[app:robustness]{Appendix~E}); the integer-dB points are for visualization only. The lower panel shows the residuals in units of the point uncertainty, with the gray band marking $\pm 2$; points beyond $\pm 6$ are drawn at the edge as open markers. (b) The same for the shots of all eight configurations pooled ($8 \times 10^8$ shots). Inset: slope $\alpha$ fitted in each configuration for the three scores, including MWPM$_{\rm uncorr}$ (values and $\chi^2_\nu$ in Table~\ref{alpha_table} of \hyperref[app:calib]{Appendix~C}); the statistical uncertainties are smaller than the markers, and the dotted line marks $\alpha = 1$.}
\label{fig:gap2}
\end{figure*}

Post-selection depends only on how a score orders the shots. Calibration asks, in addition, whether a confidence value predicts the probability that its own decoder has failed. This matters when confidence is used quantitatively, for example as soft input to an outer decoder \cite{gidney2025yoked, pattison2025hierarchical}. Expressing the confidence $g$ in dB, we fit
\begin{align}\label{true_gap_dB}
    P({\rm logical \ error}\,|\,g) = \dfrac{1}{1 + 10^{\alpha g/10}},
\end{align}
which is the one-parameter temperature scaling of the classification literature, with $\alpha = 1/T$ \cite{guo2017calibration}. If $g$ is the posterior confidence $|\Lambda|$ of a MAP decision, Eq.~(\ref{true_gap_dB}) holds with $\alpha = 1$. For a suboptimal decision rule, or a score that is only an approximation of $|\Lambda|$, no particular value of $\alpha$ is implied. A deviation from unity indicates non-ideal calibration within this one-parameter model \cite{dincua2025error}, while the goodness of fit $\chi^2_\nu$ tests whether a single rescaling adequately describes the observed failure probabilities. Importantly, calibration measures the average failure probability at a given confidence, not how accurately the score approximates the posterior for each individual syndrome.

Fig.~\ref{fig:gap2} shows the calibration of MWPM$_{\rm corr}$ and the GNN, both at $(9, 9)$ and pooled over all configurations.  Although the pooled fits have similar slopes, their large $\chi^2_\nu$ reflects differences between configurations, making the individual fits in Table~\ref{alpha_table} of \hyperref[app:calib]{Appendix~C} the more informative comparison. The most striking result is the effect of correlations on matching confidence. The uncorrelated gap is substantially miscalibrated, with $\alpha=0.825$--$0.864$ and poor agreement with a single-temperature model. Incorporating correlations brings its slope much closer to unity ($\alpha=0.941$--$0.968$), and at $d \geq 7$ the calibration is well described by a single rescaling ($\chi^2_\nu=1.4$--$3.9$). The GNN slopes also lie near unity ($\alpha=0.920$--$1.035$), but their goodness of fit varies ($\chi^2_\nu=1.2$--$45$). Thus, much of the apparent miscalibration of the standard matching gap arises from the correlations omitted by its error model, rather than from the complementary-gap construction itself.

The two scores also exhibit different calibration behavior across configurations. The correlated gap's slope varies smoothly with distance and only weakly with the number of rounds, while its local calibration slope is nearly constant at $d \geq 7$ (\hyperref[app:robustness]{Appendix~E}). The GNN slope instead varies substantially between configurations, consistent with the networks having been trained separately. Its calibration also exhibits confidence-dependent deviations from a single-temperature model, which are substantially reduced by allowing an additional offset (Platt scaling, \hyperref[app:calib]{Appendix~C}). At $d = 5$, the correlated gap likewise retains significant residual structure that an offset does not remove.

The training distribution provides a possible explanation for part of the GNN's behavior. Since the physical error rate $p$ is not supplied to the network, its target is a mixture posterior rather than the posterior at any fixed $p$. Consistent with this distinction, the GNN calibration slope decreases as $p$ increases over the tested range, with the strongest variation at $d = 5$ and the weakest at $d = 9$ (\hyperref[app:robustness]{Appendix~E}).
Its overconfidence near the upper edge of the training range  ($\alpha < 1$ at $p = 5 \times 10^{-3}$ in seven of the eight configurations) also agrees with the exact training-mixture comparison at $d = 3$ where $\Lambda_{\rm mix}$ requires $a = 0.90$. The weaker dependence at larger distances might reflect the greater information about $p$ contained in larger syndromes, although we have not tested this directly. Both matching gaps also exhibit a systematic dependence on $p$, smaller than that of the GNN at $d = 5$ and $7$, despite using weights computed at the correct error rate, indicating that the mixture alone cannot explain all calibration trends. Separating these effects would require further analysis, including controlled retraining with $p$ supplied as an input.

Finally, previous empirical calibrations of the matching gap in the surface code found $\alpha \sim 0.927$ \cite{akahoshi2025runtime} and $\alpha \sim 0.9$ \cite{gidney2025yoked}, for different noise models and gap constructions. Together with our results, this reinforces that the numerical value of $\alpha$ is setting-dependent rather than universal.

A calibration slope close to unity, however, does not guarantee that a score accurately approximates the posterior for individual syndromes. To examine this, we used the exact posterior at $d = 3$ (Fig.~\ref{fig:exact_d3}). At larger distances, where the posterior cannot be computed, we can still compare the scores through their native log-loss. For each shot, we evaluate $-\log_2 q_D(L | s)$, where $q_D(L | s)$ is the probability assigned to the observed logical outcome by the signed native confidence of decoder $D$. The expected log-loss satisfies
\begin{align}\label{eq:logloss}
    \mathbb{E}\left[-\log_2 q_D(L | s)\right] &= H(L | S) \nonumber\\
    &\quad + \mathbb{E}_s\left[D_{\rm KL}\big(P(L | s) \,\|\, q_D(L | s)\big)\right],
\end{align}
where $H(L | S)$ is the conditional entropy of the logical outcome given the syndrome at the benchmark error rate. This term depends only on the physical noise model and is therefore common to all decoders. Consequently, differences in their expected log-loss equal differences in their mean KL divergence from the exact posterior. This allows us to compare posterior approximation quality even when the posterior itself is inaccessible, using the individual predictions on every sampled shot rather than averages over confidence bins. 

The log-loss comparison provides a stronger test. At $d \geq 7$, the correlated gap achieves lower native log-loss than the GNN in every configuration (Table~\ref{alpha_table}, \hyperref[app:calib]{Appendix~C}), whereas at $d = 5$ the GNN retains its advantage. At $(9, 9)$, for example, the respective log-losses are $12.7$ and $17.0$ mbit per shot. By Eq.~(\ref{eq:logloss}), this corresponds to an estimated $4.2$ mbit per shot advantage in mean KL divergence from the exact posterior. Incorporating correlations also substantially reduces the divergence of the matching gap itself. These differences are far larger than the statistical uncertainties and are barely affected by a global rescaling. Together with the exact $d = 3$ results, they reveal a reversal: the GNN closely approximates the posterior at small distance, but the correlated gap provides the better probabilistic predictions at larger distances.

\section{Discussion and conclusion}

We have compared learned and matching-based confidence on identical rotated-surface-code syndromes, using correlation ablation, exchanged scores, calibration, and exact posterior calculations. The central finding is that the quality of soft information depends strongly on the statistical information represented by the decoder. While the GNN outperforms standard uncorrelated matching for $\kappa \gtrsim 0.5$, incorporating correlations between $X$- and $Z$-type detection events substantially improves the matching gap's discrimination and calibration, reversing the post-selection comparison at $d\geq7$. Much of the apparent superiority of learned confidence over the standard matching gap therefore arises from an incomplete matching error model, rather than from an intrinsic limitation of complementary-gap confidence.

Our analysis also reveals why comparing confidence scores requires more than post-selected error rates or calibration slopes. Exchanging scores while fixing the hard decisions shows that confidence discrimination is decoder-relative. At larger distances, each score best identifies the failures of its own decoder. Moreover, the disagreements between the decoders contain additional predictive information, suggesting that their soft outputs may be complementary rather than merely competing. By combining exact posterior calculations at $d = 3$ with log-loss comparisons at larger distances, we further distinguish confidence ranking from posterior approximation. The GNN closely reproduces the posterior of its training mixture and achieves nearly MAP-optimal decisions at $d = 3$. Yet at $d \geq 7$, the correlated matching gap has lower mean log-loss and therefore lower estimated mean divergence from the exact posterior. Excellent soft-output performance at small distance does not necessarily persist as the code grows.

These findings also highlight the promise of learned confidence. The GNN produces a decision and a meaningful soft output in a single inference pass, without explicitly constructing a DEM-based comparison between logical sectors. It retains an advantage in post-selection at $d = 5$, and its close agreement with the training-mixture posterior at $d = 3$ demonstrates that the learned logit can represent substantially more than an arbitrary failure-ranking score. This approach may be particularly valuable for complex noise models, codes without a natural matching decoder, including general qLDPC codes, and systems with many logical sectors where explicit sector comparisons become expensive \cite{lee2026efficient}. The decoder disagreements and unshared failures identified here also motivate hybrid confidence models that exploit information neither decoder captures alone. More generally, our results show that the reliability of soft information cannot be separated from the decoder's decisions or the noise information represented in its model. Whether these findings extend to other codes and noise regimes, and whether hybrid or noise-conditioned confidence models can exploit them, remain open questions for future work.

\bibliographystyle{unsrt}
\bibliography{gnn_bib}

@article{knill2005quantum,
  title={Quantum computing with realistically noisy devices},
  author={Knill, Emanuel},
  journal={Nature},
  volume={434},
  number={7029},
  pages={39--44},
  year={2005},
  publisher={Nature Publishing Group UK London}
}

@article{aliferis2007accuracy,
  title={Accuracy threshold for postselected quantum computation},
  author={Aliferis, Panos and Gottesman, Daniel and Preskill, John},
  journal={Quantum Information and Computation},
  volume={8},
  number={3--4},
  pages={181--244},
  year={2008},
  doi={10.26421/QIC8.3-4-1}
}

@article{chamberland2020very,
  title={Very low overhead fault-tolerant magic state preparation using redundant ancilla encoding and flag qubits},
  author={Chamberland, Christopher and Noh, Kyungjoo},
  journal={npj Quantum Information},
  volume={6},
  number={1},
  pages={91},
  year={2020},
  publisher={Nature Publishing Group UK London}
}

@article{li2015magic,
  title={A magic state’s fidelity can be superior to the operations that created it},
  author={Li, Ying},
  journal={New Journal of Physics},
  volume={17},
  number={2},
  pages={023037},
  year={2015},
  publisher={IOP Publishing}
}

@article{gidney2023cleaner,
  title={Cleaner magic states with hook injection},
  author={Gidney, Craig},
  journal={arXiv preprint arXiv:2302.12292},
  year={2023}
}

@article{singh2022high,
  title={High-fidelity magic-state preparation with a biased-noise architecture},
  author={Singh, Shraddha and Darmawan, Andrew S and Brown, Benjamin J and Puri, Shruti},
  journal={Physical Review A},
  volume={105},
  number={5},
  pages={052410},
  year={2022},
  publisher={APS}
}

@article{bravyi2012magic,
  title={Magic-state distillation with low overhead},
  author={Bravyi, Sergey and Haah, Jeongwan},
  journal={Physical Review A},
  volume={86},
  number={5},
  pages={052329},
  year={2012},
  publisher={APS}
}

@article{litinski2019magic,
  title={Magic state distillation: Not as costly as you think},
  author={Litinski, Daniel},
  journal={Quantum},
  volume={3},
  pages={205},
  year={2019},
  publisher={Verein zur F{\"o}rderung des Open Access Publizierens in den Quantenwissenschaften}
}

@article{lee2025low,
  title={Low-overhead magic state distillation with color codes},
  author={Lee, Seok-Hyung and Thomsen, Felix and Fazio, Nicholas and Brown, Benjamin J and Bartlett, Stephen D},
  journal={PRX Quantum},
  volume={6},
  number={3},
  pages={030317},
  year={2025},
  publisher={APS}
}

@article{gidney2024magic,
  title={Magic state cultivation: growing {T} states as cheap as {CNOT} gates},
  author={Gidney, Craig and Shutty, Noah and Jones, Cody},
  journal={arXiv preprint arXiv:2409.17595},
  year={2024}
}

@article{rosenfeld2025magic,
  title={Magic state cultivation on a superconducting quantum processor},
  author={Rosenfeld, Emma and Gidney, Craig and Roberts, Gabrielle and Morvan, Alexis and Lacroix, Nathan and Kafri, Dvir and Marshall, Jeffrey and Li, Ming and Sivak, Volodymyr and Abanin, Dmitry and others},
  journal={Science},
  pages={eaee7961},
  year={2026},
  doi={10.1126/science.aee7961}
}

@article{vaknin2026high,
  title={High rate magic state cultivation on the surface code},
  author={Vaknin, Yotam and Jacoby, Shoham and Grimsmo, Arne and Retzker, Alex},
  journal={PRX Quantum},
  volume={7},
  number={1},
  pages={010353},
  year={2026},
  publisher={APS}
}

@article{sahay2025fold,
  title={Fold-transversal surface code cultivation},
  author={Sahay, Kaavya and Tsai, Pei-Kai and Chang, Kathleen and Su, Qile and Smith, Thomas B and Singh, Shraddha and Puri, Shruti},
  journal={PRX Quantum},
  volume={7},
  number={3},
  pages={033006},
  year={2026},
  publisher={APS},
  doi={10.1103/gpvl-lg4c}
}

@article{chen2026efficient,
  title={Efficient Magic State Cultivation on {$\mathbb{RP}^2$}},
  author={Chen, Zi-Han and Chen, Ming-Cheng and Lu, Chao-Yang and Pan, Jian-Wei},
  journal={PRX Quantum},
  volume={7},
  number={1},
  pages={010315},
  year={2026},
  publisher={APS}
}

@article{claes2025cultivating,
  title={Cultivating {T} states on the surface code with only two-qubit gates},
  author={Claes, Jahan},
  journal={arXiv preprint arXiv:2509.05232},
  year={2025}
}

@article{prabhu2024distance,
  title={Distance-four quantum codes with combined postselection and error correction},
  author={Prabhu, Prithviraj and Reichardt, Ben W},
  journal={Physical Review A},
  volume={110},
  number={1},
  pages={012419},
  year={2024},
  publisher={APS}
}

@article{dobbs2025advantage,
  title={Advantage in distributed quantum computing with slow interconnects},
  author={Dobbs, Evan E and Delfosse, Nicolas and Brodutch, Aharon},
  journal={arXiv preprint arXiv:2512.10693},
  year={2025}
}

@article{akahoshi2025runtime,
  title={Runtime reduction in lattice surgery utilizing timelike soft information},
  author={Akahoshi, Yutaro and Toshio, Riki and Fujisaki, Jun and Oshima, Hirotaka and Sato, Shintaro and Fujii, Keisuke},
  journal={Physical Review Research},
  volume={8},
  number={3},
  pages={033085},
  year={2026},
  publisher={APS},
  doi={10.1103/wz3p-sgdq}
}

@article{chen2025scalable,
  title={Scalable accuracy gains from postselection in quantum error correcting codes},
  author={Chen, Hongkun and Xu, Daohong and Sommers, Grace M and Huse, David A and Thompson, Jeff D and Gopalakrishnan, Sarang},
  journal={Physical Review A},
  volume={114},
  number={3},
  pages={032455},
  year={2026},
  publisher={APS},
  doi={10.1103/mfyk-5j3x}
}

@article{bhambay2026adaptive,
  title={Adaptive Aborting Schemes for Quantum Error Correction Decoding},
  author={Bhambay, Sanidhay and Murali, Prakash and Walton, Neil and Vasantam, Thirupathaiah},
  journal={arXiv preprint arXiv:2602.16929},
  year={2026}
}

@article{bombin2024fault,
  title={Fault-tolerant postselection for low-overhead magic state preparation},
  author={Bomb{\'\i}n, H{\'e}ctor and Pant, Mihir and Roberts, Sam and Seetharam, Karthik I},
  journal={PRX Quantum},
  volume={5},
  number={1},
  pages={010302},
  year={2024},
  publisher={APS}
}

@article{poulin2006optimal,
  title={Optimal and efficient decoding of concatenated quantum block codes},
  author={Poulin, David},
  journal={Physical Review A},
  volume={74},
  number={5},
  pages={052333},
  year={2006},
  publisher={APS}
}

@article{gidney2025yoked,
  title={Yoked surface codes},
  author={Gidney, Craig and Newman, Michael and Brooks, Peter and Jones, Cody},
  journal={Nature Communications},
  volume={16},
  number={1},
  pages={4498},
  year={2025},
  publisher={Nature Publishing Group UK London}
}

@article{meister2024efficient,
  title={Efficient soft-output decoders for the surface code},
  author={Meister, Nadine and Pattison, Christopher A and Preskill, John},
  journal={arXiv preprint arXiv:2405.07433},
  year={2024}
}

@article{torlai2017neural,
  title={Neural decoder for topological codes},
  author={Torlai, Giacomo and Melko, Roger G},
  journal={Physical Review Letters},
  volume={119},
  number={3},
  pages={030501},
  year={2017},
  publisher={APS}
}

@article{krastanov2017deep,
  title={Deep neural network probabilistic decoder for stabilizer codes},
  author={Krastanov, Stefan and Jiang, Liang},
  journal={Scientific Reports},
  volume={7},
  number={1},
  pages={11003},
  year={2017},
  publisher={Nature Publishing Group UK London}
}

@article{baireuther2019neural,
  title={Neural network decoder for topological color codes with circuit level noise},
  author={Baireuther, Paul and Caio, Marcello D and Criger, Ben and Beenakker, Carlo WJ and O’Brien, Thomas E},
  journal={New Journal of Physics},
  volume={21},
  number={1},
  pages={013003},
  year={2019},
  publisher={IOP Publishing}
}

@article{meinerz2022scalable,
  title={Scalable neural decoder for topological surface codes},
  author={Meinerz, Kai and Park, Chae-Yeun and Trebst, Simon},
  journal={Physical Review Letters},
  volume={128},
  number={8},
  pages={080505},
  year={2022},
  publisher={APS}
}

@article{bausch2024learning,
  title={Learning high-accuracy error decoding for quantum processors},
  author={Bausch, Johannes and Senior, Andrew W and Heras, Francisco JH and Edlich, Thomas and Davies, Alex and Newman, Michael and Jones, Cody and Satzinger, Kevin and Niu, Murphy Yuezhen and Blackwell, Sam and others},
  journal={Nature},
  volume={635},
  number={8040},
  pages={834--840},
  year={2024},
  publisher={Nature Publishing Group UK London}
}

@article{senior2025scalable,
  title={A scalable and real-time neural decoder for topological quantum codes},
  author={Senior, Andrew W and Edlich, Thomas and Heras, Francisco JH and Zhang, Lei M and Higgott, Oscar and Spencer, James S and Applebaum, Taylor and Blackwell, Sam and Ledford, Justin and {\v{Z}}emgulyt{\.e}, Akvil{\.e} and others},
  journal={arXiv preprint arXiv:2512.07737},
  year={2025}
}

@article{gu2026scalable,
  title={Scalable Neural Decoders for Practical Fault-Tolerant Quantum Computation},
  author={Gu, Andi and Bonilla Ataides, J. Pablo and Lukin, Mikhail D and Yelin, Susanne F},
  journal={arXiv preprint arXiv:2604.08358},
  year={2026}
}

@article{lange2025data,
  title={Data-driven decoding of quantum error correcting codes using graph neural networks},
  author={Lange, Moritz and Havstr{\"o}m, Pontus and Srivastava, Basudha and Bengtsson, Isak and Bergentall, Valdemar and Hammar, Karl and Heuts, Olivia and van Nieuwenburg, Evert and Granath, Mats},
  journal={Physical Review Research},
  volume={7},
  number={2},
  pages={023181},
  year={2025},
  publisher={APS}
}

@article{lee2026efficient,
  title={Efficient post-selection for general quantum {LDPC} codes},
  author={Lee, Seok-Hyung and English, Lucas H and Bartlett, Stephen D},
  journal={npj Quantum Information},
  volume={12},
  number={1},
  pages={96},
  year={2026},
  publisher={Nature Publishing Group UK London},
  doi={10.1038/s41534-026-01242-x}
}

@article{gidney2021stim,
  doi = {10.22331/q-2021-07-06-497},
  url = {https://doi.org/10.22331/q-2021-07-06-497},
  title = {Stim: a fast stabilizer circuit simulator},
  author = {Gidney, Craig},
  journal = {{Quantum}},
  issn = {2521-327X},
  publisher = {{Verein zur F{\"{o}}rderung des Open Access Publizierens
                in den Quantenwissenschaften}},
  volume = 5,
  pages = 497,
  month = jul,
  year = 2021
}

@misc{pymatchingv2,
  author = {Higgott, Oscar and Gidney, Craig},
  title = {{PyMatching} v2},
  year = {2022},
  publisher = {GitHub},
  journal = {GitHub repository},
  howpublished = {\url{https://github.com/oscarhiggott/PyMatching}}
}

@article{haruna2025hierarchical,
  title={Hierarchical Quantum Error Correction with Hypergraph Product Code and Rotated Surface Code},
  author={Haruna, Junichi and Fujii, Keisuke},
  journal={Progress of Theoretical and Experimental Physics},
  volume={2025},
  number={10},
  pages={103A03},
  year={2025},
  publisher={Oxford University Press}
}

@article{higgott2021pymatching,
  title={{PyMatching}: A {Python} package for decoding quantum codes with minimum-weight perfect matching},
  author={Higgott, Oscar},
  journal={ACM Transactions on Quantum Computing},
  volume={3},
  number={3},
  pages={1--16},
  year={2022},
  publisher={ACM},
  doi={10.1145/3505637}
}

@article{dincua2025error,
  title={Error mitigation for logical circuits using decoder confidence},
  author={Dinc{\u{a}}, Maria and Chan, Tim and Benjamin, Simon C},
  journal={arXiv preprint arXiv:2512.15689},
  year={2025}
}

@article{wills2026forced,
  title={Forced Gap Post-Selection for Quantum {LDPC} Codes and their Operations},
  author={Wills, Adam and Yoder, Theodore J and Chuang, Isaac},
  journal={arXiv preprint arXiv:2605.20346},
  year={2026}
}

@article{bravyi2005universal,
  title={Universal quantum computation with ideal {Clifford} gates and noisy ancillas},
  author={Bravyi, Sergey and Kitaev, Alexei},
  journal={Physical Review A},
  volume={71},
  number={2},
  pages={022316},
  year={2005},
  publisher={APS}
}

@article{kishi2026even,
  title={Even More Efficient Soft-Output Decoding with Extra-Cluster Growth and Early Stopping},
  author={Kishi, Kaito and Toshio, Riki and Fujisaki, Jun and Oshima, Hirotaka and Sato, Shintaro and Fujii, Keisuke},
  journal={arXiv preprint arXiv:2602.03336},
  year={2026}
}

@article{zhong2024advantage,
  title={Advantage of quantum neural networks as quantum information decoders},
  author={Zhong, Weishun and Shtanko, Oles and Movassagh, Ramis},
  journal={arXiv preprint arXiv:2401.06300},
  year={2024}
}

@article{staples2026scalable,
  title={Scalable postselection of quantum resources},
  author={Staples, J Wilson and Fu, Winston and Thompson, Jeff D},
  journal={arXiv preprint arXiv:2603.08697},
  year={2026}
}

@inproceedings{guo2017calibration,
  title={On calibration of modern neural networks},
  author={Guo, Chuan and Pleiss, Geoff and Sun, Yu and Weinberger, Kilian Q},
  booktitle={Proceedings of the 34th International Conference on Machine Learning},
  series={Proceedings of Machine Learning Research},
  volume={70},
  pages={1321--1330},
  year={2017},
  organization={PMLR}
}

@article{xu2026diffqec,
  title={{DiffQEC}: A versatile diffusion model for quantum error correction},
  author={Xu, Tianyi and Liu, Qinglong and Wang, Maolin and Zhang, Fei and Zhao, Zhe and Wang, Yang and Wei, Ye},
  journal={arXiv preprint arXiv:2604.24640},
  year={2026}
}

@article{seip2026machine,
  title={Machine Learning Optimal Quantum Error Correction Thresholds},
  author={Seip, Dominik and Colmenarez, Luis and Schmitt, Markus and M{\"u}ller, Markus},
  journal={arXiv preprint arXiv:2606.22194},
  year={2026}
}

@article{haug2026machine,
  title={Machine-learned syndrome post-selection for reliable quantum error correction},
  author={Haug, Tobias and Canabarro, Askery and Aolita, Leandro},
  journal={arXiv preprint arXiv:2607.19563},
  year={2026}
}

@article{hernando2026correlated,
  title={Correlated decoding loses the surface code's effective distance below threshold},
  author={Hernando de la Fuente, C{\'e}sar and Majaniemi, Joonas and van Nieuwenburg, Evert and Frohnert, Felix},
  journal={arXiv preprint arXiv:2610.26785},
  year={2026}
}

@article{fowler2013optimal,
  title={Optimal complexity correction of correlated errors in the surface code},
  author={Fowler, Austin G.},
  journal={arXiv preprint arXiv:1310.0863},
  year={2013}
}

@article{pattison2025hierarchical,
  title={Hierarchical memories: Simulating quantum {LDPC} codes with local gates},
  author={Pattison, Christopher A. and Krishna, Anirudh and Preskill, John},
  journal={Quantum},
  volume={9},
  pages={1728},
  year={2025},
  doi={10.22331/q-2025-05-05-1728}
}

@misc{pymatchingrs,
  author = {Zhang, Yiming},
  title = {{pymatching-rs}: {Rust} bindings for {PyMatching}},
  year = {2026},
  publisher = {GitHub},
  journal = {GitHub repository},
  howpublished = {\url{https://github.com/inmzhang/pymatching-rs}}
}

@misc{dentelski2026code,
  author = {Dentelski, David},
  title = {{GNN-confidence-score}},
  year = {2026},
  publisher = {GitHub},
  journal = {GitHub repository},
  howpublished = {\url{https://github.com/DavidDentelski/GNN-confidence-score}}
}

@article{delong1988comparing,
  title={Comparing the areas under two or more correlated receiver operating characteristic curves: a nonparametric approach},
  author={DeLong, Elizabeth R. and DeLong, David M. and Clarke-Pearson, Daniel L.},
  journal={Biometrics},
  volume={44},
  number={3},
  pages={837--845},
  year={1988},
  doi={10.2307/2531595}
}

\clearpage
\onecolumngrid

\section{Appendix A: Statistical conventions and failure counts}
\phantomsection\label{app:stats}

Unless otherwise stated, interval estimates are $95\%$ intervals. The bands on the post-selection curves (Figs.~\ref{fig:LER} and \ref{fig:crossed}) are normal intervals, $\pm 1.96$ binomial standard errors of the post-selected LER, and the calibration points of Figs.~\ref{fig:gap2} and \ref{fig:calib_supp} carry Wilson score intervals.

Two tie conventions are used throughout. First, when the acceptance threshold at a given $\kappa$ falls inside a block of shots that share a single confidence value, the block's failures are accepted in proportion to the accepted fraction of the block, so that the post-selected LER is the average over random orderings of the tied shots. This matters wherever tie blocks are macroscopic, most notably for the trivial syndromes and for the block of shots at the bound of the uncorrelated gap at $d = 5$ (see the discussion near Fig.~\ref{fig:gap1}). Second, AUC values use midranks: a tie between a failure and a correctly decoded shot receives half credit. This equals the AUC averaged over random orderings of the tied shots, consistent with the fractional tie rule of the post-selection curves. The AUC summarizes the ranking over all acceptance rates and is invariant under strictly monotonic transformations of the score, but it does not determine the post-selected LER at a particular $\kappa$, which is why it is reported together with Figs.~\ref{fig:LER} and \ref{fig:crossed}.

Since all decoders are evaluated on the same $10^8$ shots per configuration, ratios of post-selected LERs, and the ratios $R_{\rm pair}/R_{\rm GNN}$ of the selection gains in Fig.~\ref{fig:LER}, are estimated with a paired resampling scheme: at each $\kappa$, every shot is classified into the joint categories defined by its (accepted, failed) status under each decoder-confidence combination, the category counts are resampled from a multinomial distribution, and the ratio is recomputed in every replicate under the same fractional tie rule (for the selection gains, with the hard-decision LERs $L_D(1)$ of the same replicate). Paired uncertainties of AUC differences are computed with DeLong's method \cite{delong1988comparing}, which accounts for the evaluation of both scores on the same shots.

A comparison is claimed in the text only where the corresponding paired $95\%$ interval excludes unity (for LER ratios) or zero (for AUC differences). In Fig.~\ref{fig:LER}, a point is displayed only if it rests on at least $20$ accepted failures (for the ratios, in both pairs), and in Fig.~\ref{fig:crossed} only if all four curves of the panel do, which keeps the relative $95\%$ uncertainty of each displayed point below about $45\%$. The accepted-failure counts behind the points at $p = 5 \times 10^{-3}$ are listed in Table~\ref{counts_table} for the native pairs of Fig.~\ref{fig:LER} and in Table~\ref{crossed_counts_table} for the exchanged pairs of Fig.~\ref{fig:crossed}.

\begin{table}[H]
\centering
\begin{tabular}{ll| c c c c c c c c c}
\toprule
 & & \multicolumn{9}{c}{Accepted failures at acceptance rate $\kappa$} \\
Config & Pair & 0.2 & 0.3 & 0.4 & 0.5 & 0.6 & 0.7 & 0.8 & 0.9 & 1.0 \\
\hline
\midrule
$(5, 5)$ & MWPM$_{\rm uncorr}$ + $g_{\rm uncorr}$ & 70 & 253 & 538 & 1116 & 2380 & 7482 & 28021 & 137994 & 1407388 \\
$(5, 5)$ & MWPM$_{\rm corr}$ + $g_{\rm corr}$ & 12 & 64 & 158 & 388 & 987 & 3093 & 11025 & 61956 & 1088447 \\
$(5, 5)$ & GNN + $g_{\rm GNN}$ & 20 & 60 & 140 & 342 & 756 & 2013 & 7153 & 41724 & 938002 \\
\hline
$(7, 7)$ & MWPM$_{\rm uncorr}$ + $g_{\rm uncorr}$ & 9 & 43 & 172 & 431 & 1063 & 2906 & 9671 & 46229 & 1000860 \\
$(7, 7)$ & MWPM$_{\rm corr}$ + $g_{\rm corr}$ & 2 & 8 & 22 & 62 & 207 & 618 & 2105 & 11023 & 597796 \\
$(7, 7)$ & GNN + $g_{\rm GNN}$ & 12 & 27 & 78 & 178 & 423 & 1122 & 3726 & 18162 & 731541 \\
\hline
$(9, 9)$ & MWPM$_{\rm uncorr}$ + $g_{\rm uncorr}$ & 1 & 7 & 35 & 118 & 342 & 981 & 3152 & 16272 & 676322 \\
$(9, 9)$ & MWPM$_{\rm corr}$ + $g_{\rm corr}$ & 0 & 1 & 2 & 8 & 31 & 100 & 366 & 2049 & 307200 \\
$(9, 9)$ & GNN + $g_{\rm GNN}$ & 27 & 33 & 46 & 74 & 139 & 313 & 967 & 4626 & 414621 \\
\bottomrule
\end{tabular}
\caption{Accepted logical failures behind the top row of Fig.~\ref{fig:LER}, at $p = 5 \times 10^{-3}$ with $10^8$ shots per configuration. Acceptance-threshold ties are averaged uniformly (fractionally accepted), so the entries are tie-averaged expectations, rounded to integers. Points resting on fewer than $20$ failures are not shown in the figure.}
\label{counts_table}
\end{table}

\begin{table}
\centering
\begin{tabular}{ll| c c c c c c c c c}
\toprule
 & & \multicolumn{9}{c}{Accepted failures at acceptance rate $\kappa$} \\
Config & Curve & 0.2 & 0.3 & 0.4 & 0.5 & 0.6 & 0.7 & 0.8 & 0.9 & 1.0 \\
\hline
\midrule
$(5, 5)$ & MWPM$_{\rm corr}$, ranked by $g_{\rm GNN}$ & 20 & 60 & 140 & 343 & 760 & 2072 & 7843 & 51090 & 1088447 \\
$(5, 5)$ & GNN, ranked by $g_{\rm corr}$ & 14 & 70 & 181 & 438 & 1123 & 3413 & 12023 & 66807 & 938002 \\
$(7, 7)$ & MWPM$_{\rm corr}$, ranked by $g_{\rm GNN}$ & 12 & 30 & 83 & 208 & 516 & 1466 & 5177 & 27211 & 597796 \\
$(7, 7)$ & GNN, ranked by $g_{\rm corr}$ & 252 & 681 & 1470 & 2799 & 4912 & 8549 & 17107 & 49846 & 731541 \\
$(9, 9)$ & MWPM$_{\rm corr}$, ranked by $g_{\rm GNN}$ & 38 & 51 & 73 & 125 & 231 & 555 & 1862 & 9164 & 307200 \\
$(9, 9)$ & GNN, ranked by $g_{\rm corr}$ & 123 & 253 & 465 & 833 & 1461 & 2738 & 6126 & 19882 & 414621 \\
\hline
$(5, 5)$ & MWPM$_{\rm uncorr}$, ranked by $g_{\rm GNN}$ & 20 & 60 & 146 & 372 & 915 & 2818 & 12575 & 85413 & 1407388 \\
$(5, 5)$ & GNN, ranked by $g_{\rm uncorr}$ & 74 & 264 & 568 & 1171 & 2486 & 7763 & 28750 & 134608 & 938002 \\
$(7, 7)$ & MWPM$_{\rm uncorr}$, ranked by $g_{\rm GNN}$ & 22 & 67 & 195 & 509 & 1334 & 4031 & 14254 & 68399 & 1000860 \\
$(7, 7)$ & GNN, ranked by $g_{\rm uncorr}$ & 195 & 630 & 1585 & 2907 & 5182 & 10157 & 23596 & 79619 & 731541 \\
$(9, 9)$ & MWPM$_{\rm uncorr}$, ranked by $g_{\rm GNN}$ & 251 & 348 & 564 & 1013 & 2052 & 4785 & 13537 & 50820 & 676322 \\
$(9, 9)$ & GNN, ranked by $g_{\rm uncorr}$ & 103 & 226 & 477 & 873 & 1675 & 3547 & 9041 & 33826 & 414621 \\
\bottomrule
\end{tabular}
\caption{Accepted logical failures behind the exchanged curves of Fig.~\ref{fig:crossed}, at $p = 5 \times 10^{-3}$ with $10^8$ shots per configuration (tie-averaged expectations, rounded to integers). The native curves are those of Table~\ref{counts_table}.}
\label{crossed_counts_table}
\end{table}

\section{Appendix B: Correlation-aware complementary gap}
\phantomsection\label{app:corr}

\emph{Implementation and validation.} The correlated gap follows the two-pass reweighting of \texttt{PyMatching} 2.4 described in the main text after Eq.~(\ref{MWPM_gap}). After the first pass, the implied weights of the correlated partners of all first-pass edges are applied to the matching graph, and the two logical sectors are then matched on this single reweighted graph, with the virtual observable detector forced to $0$ and to $1$; finally the reweighting is undone. We use two implementations with identical results: a reference implementation that applies the reweighting through the \texttt{PyMatching} Python interface, and a faster one that performs the same steps inside the \texttt{PyMatching} C++ core through a small patch \cite{dentelski2026code}. By construction, the lighter of the two forced matchings is the correlated second-pass matching of \texttt{PyMatching}. The decision coincides on all $10^8$ shots of each of the eight configurations except exact ties ($g_{\rm corr} = 0$; between $36$ shots at $(9, 9)$ and $1907$ at $(5, 11)$). On all $10^8$ shots of the $(5, 5)$ and $(7, 7)$ configurations, the weight of the decided sector also agrees with \texttt{PyMatching}'s correlated matching weight to numerical precision.

\emph{Per-sector reweighting.} A different construction reweights each sector from its own first pass, i.e., the complementary sector is matched on a graph reweighted by a first pass that is itself forced into that sector \cite{pymatchingrs}. The two sectors are then compared on different graphs. On the same shots and with the same decisions, this per-sector gap is a much weaker confidence score: the AUC for ranking MWPM's failures drops from $0.9753$ to $0.9609$ at $d = 5$ and from $0.9861$ to $0.9741$ at $d = 7$, and the post-selected LER at $\kappa = 0.9$ increases by a factor of $1.84$ ($95\%$ interval $1.82$--$1.85$, bootstrap over blocks of $10^4$ shots) at $d = 5$ and $2.80$ ($2.76$--$2.85$) at $d = 7$. At $d = 7$ this would place the matching pair behind the GNN pair at $\kappa = 0.9$ on the same shots ($3.4 \times 10^{-4}$ versus $2.0 \times 10^{-4}$), the opposite of the conclusion obtained with the shared graph. On about $1\%$ of the shots ($0.97\%$ and $0.64\%$), the complementary sector has smaller minimum weight than the decided one after its own reweighting, a sign inconsistency hidden by the absolute value, and $21$--$22\%$ of these shots are failures. A correlated complementary gap is therefore not uniquely defined until its construction is specified. For the shared-reweighting approach used here, comparing the two sectors on a common graph, as in Ref.~\cite{gidney2025yoked}, is the consistent construction, and the choice materially affects the conclusions.

\section{Appendix C: Calibration of all configurations}
\phantomsection\label{app:calib}

\begin{figure*}
    \centering
    \includegraphics[width=0.75\textwidth]{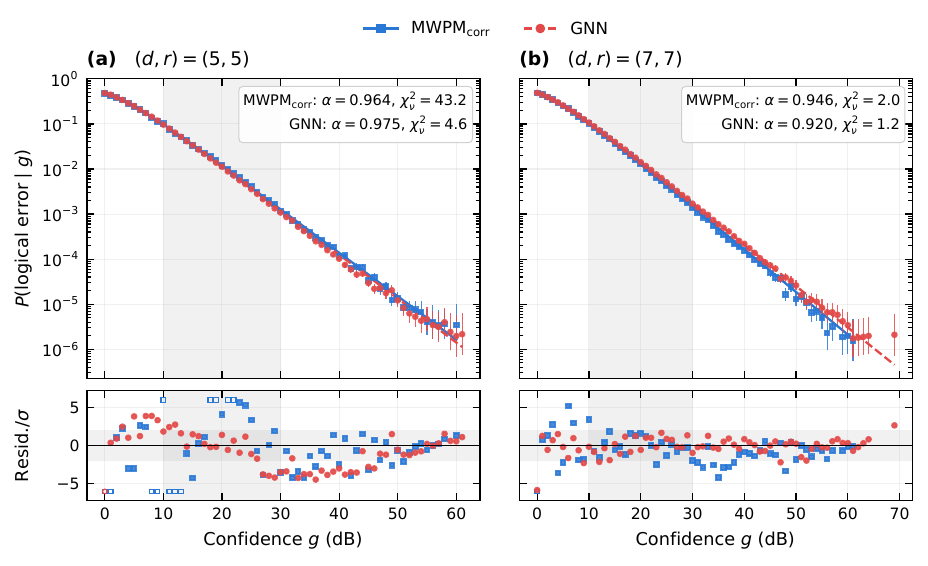}
    \caption{Empirical calibration of MWPM$_{\rm corr}$ and the GNN, with maximum-likelihood fits of Eq.~(\ref{true_gap_dB}), at (a) $(d, r) = (5, 5)$ and (b) $(d, r) = (7, 7)$, at $p = 5 \times 10^{-3}$; conventions as in Fig.~\ref{fig:gap2}(a).}
\label{fig:calib_supp}
\end{figure*}

Fig.~\ref{fig:calib_supp} shows the calibration curves at $(5, 5)$ and $(7, 7)$, completing Fig.~\ref{fig:gap2}(a), and Table~\ref{alpha_table} collects the fits of all configurations for the three scores: the one-parameter slope $\alpha$ and its goodness of fit, a two-parameter fit with an additional offset $c$, $P = 1/[1 + 10^{(\alpha_2 g - c)/10}]$, for MWPM$_{\rm corr}$ and the GNN, and the native log-loss. The statistical uncertainty of the slopes is at most $0.001$ (Fisher and binomial bootstrap agree), so that the differences between configurations are highly resolved. For the correlated gap, a weighted linear model in $d$ and $r$ attributes the variation of the slopes to the distance ($-0.0079 \pm 0.0002$ per unit of $d$) and none to the number of rounds ($+0.0001 \pm 0.0001$). The GNN slopes do not follow such a model: neighboring configurations differ by up to $0.075$, i.e., by $13$--$86$ standard deviations, as expected for separately trained networks. The offset reduces the GNN's $\chi^2_\nu$ to $0.9$--$4.8$, with fitted offsets between $-0.66$ and $+0.22$ dB, while it barely changes the correlated gap, whose residual structure at $d = 5$ is therefore not a simple shift. A global rescaling of the scores lowers the native log-loss by at most $0.21$ mbit per shot for MWPM$_{\rm corr}$ and the GNN, and by at most $2.41$ mbit for MWPM$_{\rm uncorr}$.

\begin{table*}
\centering
\begin{tabular}{ll|cccccccc|c|c}
\toprule
 & & $(5, 5)$ & $(5, 7)$ & $(5, 9)$ & $(5, 11)$ & $(7, 7)$ & $(7, 9)$ & $(7, 11)$ & $(9, 9)$ & Pooled & IVW mean \\
\hline
\midrule
MWPM$_{\rm corr}$ & $\alpha$ & 0.964 & 0.968 & 0.967 & 0.965 & 0.946 & 0.949 & 0.949 & 0.941 & 0.961 & 0.960 \\
 & $\chi^2_\nu$ & 43.2 & 40.7 & 45.4 & 55.0 & 2.0 & 3.5 & 3.9 & 1.4 & 149.2 & -- \\
 & $\alpha_2$ (2-par.) & 0.935 & 0.946 & 0.947 & 0.951 & 0.948 & 0.952 & 0.953 & 0.948 & -- & -- \\
 & $\chi^2_\nu$ (2-par.) & 34.9 & 35.7 & 41.8 & 55.9 & 2.0 & 3.3 & 3.6 & 1.3 & -- & -- \\
 & log-loss & 44.52 & 59.88 & 75.03 & 89.93 & 24.38 & 30.93 & 37.40 & 12.72 & -- & -- \\
\hline
GNN & $\alpha$ & 0.975 & 0.966 & 0.938 & 0.991 & 0.920 & 0.960 & 0.983 & 1.035 & 0.968 & 0.967 \\
 & $\chi^2_\nu$ & 4.6 & 14.0 & 44.9 & 14.2 & 1.2 & 12.7 & 8.0 & 5.2 & 69.3 & -- \\
 & $\alpha_2$ (2-par.) & 0.988 & 0.941 & 0.899 & 0.971 & 0.914 & 0.933 & 0.964 & 1.008 & -- & -- \\
 & $\chi^2_\nu$ (2-par.) & 2.5 & 2.0 & 4.8 & 1.2 & 0.9 & 2.7 & 1.6 & 1.3 & -- & -- \\
 & log-loss & 38.25 & 51.94 & 65.49 & 78.51 & 29.55 & 37.95 & 48.36 & 16.96 & -- & -- \\
\hline
MWPM$_{\rm uncorr}$ & $\alpha$ & 0.864 & 0.855 & 0.847 & 0.839 & 0.837 & 0.832 & 0.827 & 0.825 & 0.843 & 0.842 \\
 & $\chi^2_\nu$ & 239.8 & 380.6 & 493.0 & 634.8 & 95.9 & 133.3 & 183.2 & 40.1 & 1985.9 & -- \\
 & log-loss & 59.26 & 83.29 & 106.82 & 129.62 & 41.60 & 55.08 & 68.33 & 28.36 & -- & -- \\
\bottomrule
\end{tabular}
\caption{Calibration of the three confidence scores at $p = 5 \times 10^{-3}$ for every configuration $(d, r)$, for the shot-pooled data, and as the inverse-variance-weighted (IVW) descriptive mean of the per-configuration slopes. $\alpha$: binomial maximum-likelihood slope of Eq.~(\ref{true_gap_dB}) over the pre-specified window $g \in [10, 30]$ dB (statistical uncertainty at most $0.001$, omitted); $\chi^2_\nu$: reduced $\chi^2$ over integer-dB bins in the window; $\alpha_2$, $\chi^2_\nu$ (2-par.): the same with an additional offset; log-loss: mean of $-\log_2 q(L|s)$ over all shots, with $q(L = 0|s) = \sigma(\lambda(s))$ at the native scale, in mbit per shot; by Eq.~(\ref{eq:logloss}), differences in log-loss between scores in the same configuration estimate differences in their mean KL divergence from the fixed-$p$ posterior. The configurations correspond to different trained networks and do not share a single true $\alpha$; the per-configuration values are the primary result.}
\label{alpha_table}
\end{table*}

\section{Appendix D: Shot-level structure of the confidence scores}
\phantomsection\label{app:pairing}

\begin{figure*}
\centering
\includegraphics[width=0.8\textwidth]{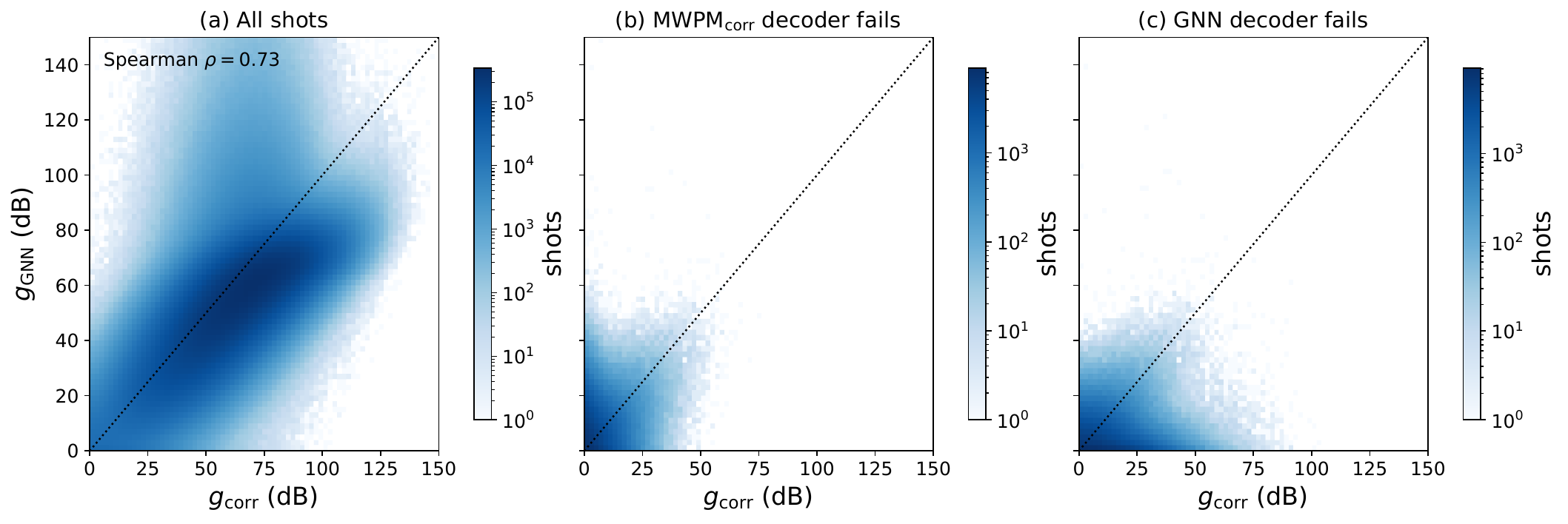}
    \caption{Joint distribution of the two confidence scores on the same shots at $(d, r) = (9, 9)$, $p = 5 \times 10^{-3}$, $10^8$ shots ($2$-dB bins, logarithmic color scale, axes truncated at $150$ dB). The dotted line marks $g_{\rm GNN} = g_{\rm corr}$. (a) All shots. (b) Shots on which MWPM$_{\rm corr}$ fails. (c) Shots on which the GNN fails.}
\label{fig:pairing}
\end{figure*}

Fig.~\ref{fig:pairing} shows the joint distribution of $g_{\rm corr}$ and $g_{\rm GNN}$ at $(9, 9)$. Over all shots the two scores are only moderately rank-correlated (Spearman $\rho = 0.73$), far from monotone transformations of one another, so the exchange of Fig.~\ref{fig:crossed} and Table~\ref{auc_table} compares genuinely different orderings. The failures of each decoder concentrate at low values of both scores, with two tails. In panel (b), the vertical tail consists of MWPM failures with a small gap but a large logit: $13\%$ of MWPM's failures carry $g_{\rm GNN} > 20$ dB, and on $58\%$ of these the GNN decodes correctly, so that the large logit expresses confidence in the opposite sector. In panel (c), the horizontal tail consists of GNN failures with a large gap: $25\%$ of the GNN's failures carry $g_{\rm corr} > 20$ dB, and MWPM decodes $81\%$ of them correctly ($9\%$ and $88\%$ for $g_{\rm corr} > 30$ dB). These tails consist largely of the decision-disagreement shots discussed with the exchange in the main text. Since the rows $|g|$ and $\pm g$ of Table~\ref{auc_table} differ only on such shots, the table quantifies their role: at $d \geq 7$, restoring the sign of the other score on these shots recovers $81$--$86\%$ of the exchange penalty in AUC for MWPM, and more than the whole penalty for the GNN. The failure sets of the two decoders overlap only partly: the fraction of MWPM's failures shared by the GNN is $60\%$, $52\%$ and $46\%$ at $(5, 5)$, $(7, 7)$ and $(9, 9)$.

Table~\ref{mechanism_table} resolves the same-decoder AUC difference for MWPM's failures by the value of their own gap. For each bin, it lists the mean fraction of correctly decoded shots that a failure outranks under each score, and the contribution of the bin to the AUC difference, split into failures shared with the GNN and failures of MWPM only. At $d \geq 7$, the advantage of the gap comes from the failures with $g_{\rm corr} < 10$ dB, mostly from those that the GNN does not share, whereas the logit demotes failures with $g_{\rm corr} > 20$ dB more effectively at every distance. These high-gap failures are few; together with those at $10$--$20$ dB, their contribution outweighs that of the near-tied failures only at $(5, 5)$, where the two contributions nearly cancel. The table thus shows where the ranking difference arises; it does not identify the features of the syndrome to which the two scores respond differently.

\begin{table*}
\centering
\begin{tabular}{ll|rr|cc|rrr}
\toprule
 & & & & \multicolumn{2}{c|}{correct shots outranked} & \multicolumn{3}{c}{contribution to $\Delta$AUC ($\times 10^{3}$)} \\
Config & $g_{\rm corr}$ (dB) & failures & fraction & by $g_{\rm corr}$ & by $g_{\rm GNN}$ & all & shared & only \\
\hline
\midrule
$(5, 5)$ & 0--5 & 496,265 & 45.6\% & 0.4\% & 1.5\% & $+5.14$ & $+1.08$ & $+4.06$ \\
$(5, 5)$ & 5--10 & 287,569 & 26.4\% & 1.5\% & 1.5\% & $-0.00$ & $+0.03$ & $-0.03$ \\
$(5, 5)$ & 10--15 & 160,076 & 14.7\% & 3.6\% & 2.5\% & $-1.63$ & $-1.07$ & $-0.56$ \\
$(5, 5)$ & 15--20 & 82,221 & 7.6\% & 6.8\% & 4.3\% & $-1.88$ & $-1.58$ & $-0.30$ \\
$(5, 5)$ & 20--30 & 53,410 & 4.9\% & 12.7\% & 8.1\% & $-2.28$ & $-2.15$ & $-0.13$ \\
$(5, 5)$ & 30--40 & 7,713 & 0.7\% & 26.2\% & 17.3\% & $-0.63$ & $-0.63$ & $-0.00$ \\
$(5, 5)$ & 40--60 & 1,191 & 0.1\% & 47.7\% & 36.4\% & $-0.12$ & $-0.12$ & $+0.00$ \\
$(5, 5)$ & $\geq 60$ & 2 & 0.0\% & 83.6\% & 55.9\% & $-0.00$ & $-0.00$ & $+0.00$ \\
$(5, 5)$ & total & 1,088,447 & 100\% & & & $-1.41$ & $-4.44$ & $+3.03$ \\
\hline
$(7, 7)$ & 0--5 & 273,468 & 45.7\% & 0.2\% & 2.2\% & $+9.36$ & $+2.17$ & $+7.18$ \\
$(7, 7)$ & 5--10 & 155,627 & 26.0\% & 0.8\% & 2.0\% & $+3.19$ & $+1.55$ & $+1.63$ \\
$(7, 7)$ & 10--15 & 85,091 & 14.2\% & 1.8\% & 2.4\% & $+0.80$ & $+0.79$ & $+0.01$ \\
$(7, 7)$ & 15--20 & 44,412 & 7.4\% & 3.4\% & 3.3\% & $-0.08$ & $+0.19$ & $-0.27$ \\
$(7, 7)$ & 20--30 & 31,884 & 5.3\% & 6.8\% & 5.4\% & $-0.76$ & $-0.48$ & $-0.28$ \\
$(7, 7)$ & 30--40 & 6,094 & 1.0\% & 15.5\% & 11.0\% & $-0.46$ & $-0.42$ & $-0.04$ \\
$(7, 7)$ & 40--60 & 1,205 & 0.2\% & 31.9\% & 22.3\% & $-0.19$ & $-0.19$ & $-0.00$ \\
$(7, 7)$ & $\geq 60$ & 15 & 0.0\% & 72.2\% & 49.3\% & $-0.01$ & $-0.01$ & $+0.00$ \\
$(7, 7)$ & total & 597,796 & 100\% & & & $+11.84$ & $+3.62$ & $+8.22$ \\
\hline
$(9, 9)$ & 0--5 & 138,580 & 45.1\% & 0.1\% & 1.9\% & $+8.31$ & $+1.42$ & $+6.89$ \\
$(9, 9)$ & 5--10 & 79,287 & 25.8\% & 0.4\% & 1.7\% & $+3.27$ & $+1.11$ & $+2.16$ \\
$(9, 9)$ & 10--15 & 43,522 & 14.2\% & 0.9\% & 1.7\% & $+1.12$ & $+0.71$ & $+0.40$ \\
$(9, 9)$ & 15--20 & 23,240 & 7.6\% & 1.7\% & 2.1\% & $+0.27$ & $+0.35$ & $-0.07$ \\
$(9, 9)$ & 20--30 & 17,641 & 5.7\% & 3.6\% & 3.2\% & $-0.21$ & $+0.01$ & $-0.22$ \\
$(9, 9)$ & 30--40 & 3,991 & 1.3\% & 8.6\% & 6.2\% & $-0.31$ & $-0.24$ & $-0.07$ \\
$(9, 9)$ & 40--60 & 924 & 0.3\% & 20.0\% & 13.4\% & $-0.20$ & $-0.19$ & $-0.01$ \\
$(9, 9)$ & $\geq 60$ & 15 & 0.0\% & 53.5\% & 22.4\% & $-0.02$ & $-0.02$ & $+0.00$ \\
$(9, 9)$ & total & 307,200 & 100\% & & & $+12.24$ & $+3.16$ & $+9.08$ \\
\bottomrule
\end{tabular}
\caption{Anatomy of the ranking of MWPM$_{\rm corr}$'s failures at $p = 5 \times 10^{-3}$, with failures grouped by their own gap $g_{\rm corr}$: number and fraction of failures, mean fraction of correctly decoded shots that a failure outranks (i.e., that receive a lower confidence than the failure; ties receive half credit) under each score, and contribution of the bin to the AUC difference $\Delta{\rm AUC} = {\rm AUC}(g_{\rm corr}) - {\rm AUC}(|g_{\rm GNN}|)$, in total and split into failures shared with the GNN and failures of MWPM$_{\rm corr}$ only.}
\label{mechanism_table}
\end{table*}

\emph{High-confidence failures of the GNN.} At $(9, 9)$, the post-selected LER of the GNN stops decreasing below $\kappa \approx 0.4$ (Table~\ref{counts_table}). This floor is set by a small group of failures with extreme logits: $15$ of the $27$ failures accepted at $\kappa = 0.2$ have $g_{\rm GNN} = 78$--$240$ dB, whereas the calibration of the GNN, Eq.~(\ref{true_gap_dB}) with the fitted $\alpha$, predicts $0.9$ accepted failures. These are hard syndromes: MWPM$_{\rm corr}$ also fails on $35$ of the $46$ GNN failures accepted at $\kappa = 0.4$, but assigns them a median gap of $37$ dB, below its own threshold of $67$ dB, and they carry more detection events than correctly decoded shots of similar logit (median $64$ versus $49$), as do the GNN's other failures. No feature we examined distinguishes them from these other failures. The excess of high-confidence failures, consistent with the decreasing local slopes of Fig.~\ref{fig:local_slopes}, appears in most configurations, but flattens the post-selection curve only at $(9, 9)$, where the remaining tail contributes almost no failures.

\section{Appendix E: Robustness checks}
\phantomsection\label{app:robustness}

This appendix verifies that the conclusions are insensitive to the trivial-syndrome convention, to the calibration fit window, and to the physical error rate, and resolves the calibration slopes by confidence.

\begin{table*}
\centering
\begin{tabular}{l|cc|cc|cc}
\toprule
Config & trivial fraction & conf.\ (dB) & $\Delta$AUC$_{\rm MWPM}$ & $\Delta$AUC$_{\rm GNN}$ & AUC$^{\rm excl}_{g_{\rm GNN}\to{\rm MWPM}}$ & AUC$^{\rm excl}_{g_{\rm corr}\to{\rm MWPM}}$ \\
\hline
\midrule
$(5, 5)$ & $1.4\%$ & 88.22 & $-0.0014$ & $+0.0093$ & 0.9764 & 0.9749 \\
$(5, 7)$ & $0.26\%$ & 88.22 & $+0.0001$ & $+0.0128$ & 0.9678 & 0.9679 \\
$(5, 9)$ & $0.05\%$ & 88.22 & $+0.0010$ & $+0.0159$ & 0.9599 & 0.9610 \\
$(5, 11)$ & $0.0094\%$ & 88.22 & $+0.0025$ & $+0.0191$ & 0.9519 & 0.9544 \\
$(7, 7)$ & $0.00055\%$ & 125.47 & $+0.0118$ & $+0.0147$ & 0.9743 & 0.9861 \\
$(7, 9)$ & $1.8 \times 10^{-5}\%$ & 125.47 & $+0.0163$ & $+0.0158$ & 0.9663 & 0.9826 \\
$(7, 11)$ & $0$ & -- & $+0.0211$ & $+0.0236$ & 0.9582 & 0.9793 \\
$(9, 9)$ & $0$ & -- & $+0.0122$ & $+0.0135$ & 0.9799 & 0.9922 \\
\bottomrule
\end{tabular}
\caption{Robustness of the discrimination results at $p = 5 \times 10^{-3}$ to the trivial syndromes, which bypass the GNN and are assigned the decision $0$ and, as confidence, the maximal matching gap (column ``conf.''). No trivial-syndrome shot in the sampled data carries a logical error. $\Delta{\rm AUC}_D$ is the AUC of decoder $D$'s failures ranked by its own score minus that ranked by the other decoder's score (all shots); the paired standard errors are below $10^{-4}$ throughout. ${\rm AUC}^{\rm excl}$ is the tie-aware AUC of MWPM$_{\rm corr}$'s failures recomputed without the trivial shots, ranked by either score. AUC ties receive half credit throughout.}
\label{robustness_table}
\end{table*}

\emph{Trivial syndromes.} Trivial syndromes bypass the GNN; they are assigned the decision $0$ and, as confidence, the matching gap of the trivial syndrome, which is the maximal gap. We identify these shots exactly from the saved confidence arrays: since the sentinel is copied from the matching computation, their signed confidences are bit-identical and share one common value. Accidental floating-point coincidences between a matching weight and a network logit occur only at scattered unique values and are counted separately. Table~\ref{robustness_table} characterizes this population: the trivial fraction falls from $1.4\%$ at $(5, 5)$ to zero at $(7, 11)$ and $(9, 9)$, the sentinel value is independent of $r$ ($88.22$ dB at $d = 5$ and $125.47$ dB at $d = 7$), and no trivial shot in any of the eight samples carries a logical error. Their effect on the results is negligible: the residual reversal at $(5, 5)$ survives their exclusion (${\rm AUC} = 0.9764$ versus $0.9749$), and the calibration fits cannot be affected, since the sentinel lies far outside the fit window; trivial shots are nevertheless excluded from the GNN fits, as their confidence is not produced by the network.

\emph{Fit window.} The calibration analysis uses a common fitting window $g \in [10, 30]$ dB for all decoders and configurations. This focuses the fits on intermediate confidences, excluding the near-tied regime at low confidence and the high-confidence tail where failures are rare. To test sensitivity to this choice, Table~\ref{window_table} repeats the fits over five alternative windows, including the wider interval $[0, 50]$ dB. The slopes shift by at most $0.009$ for MWPM$_{\rm corr}$ and $0.005$ for the GNN, without changing the conclusions of the main text. The goodness of fit depends on the window, and is best for both scores in the narrow interval $[15, 25]$ dB.

\begin{table*}
\centering
\begin{tabular}{l|cc|cc}
\toprule
 & \multicolumn{2}{c|}{MWPM$_{\rm corr}$} & \multicolumn{2}{c}{GNN} \\
Fit window (dB) & $\alpha$ range & median $\chi^2_\nu$ & $\alpha$ range & median $\chi^2_\nu$ \\
\hline
\midrule
$[10, 30]$ (baseline) & 0.941--0.968 & 22.3 & 0.920--1.035 & 10.4 \\
$[5, 30]$ & 0.941--0.969 (0.002) & 27.0 & 0.920--1.040 (0.005) & 14.9 \\
$[10, 40]$ & 0.942--0.969 (0.002) & 16.4 & 0.920--1.031 (0.004) & 10.8 \\
$[5, 40]$ & 0.942--0.970 (0.003) & 20.5 & 0.920--1.036 (0.001) & 16.2 \\
$[15, 25]$ & 0.941--0.960 (0.009) & 3.7 & 0.918--1.032 (0.003) & 3.1 \\
$[0, 50]$ & 0.942--0.970 (0.003) & 20.1 & 0.920--1.037 (0.002) & 19.2 \\
\bottomrule
\end{tabular}
\caption{Robustness of the calibration slopes to the fit window: for each window, the range of $\alpha$ across the eight configurations, with the largest per-configuration shift relative to the baseline window $g \in [10, 30]$ dB in parentheses, and the median $\chi^2_\nu$ (binomial maximum-likelihood fits as in Table~\ref{alpha_table}; the interval $[0, 50]$ dB provides a wider-range robustness check).}
\label{window_table}
\end{table*}

\begin{figure*}
\centering
\includegraphics[width=0.7\textwidth]{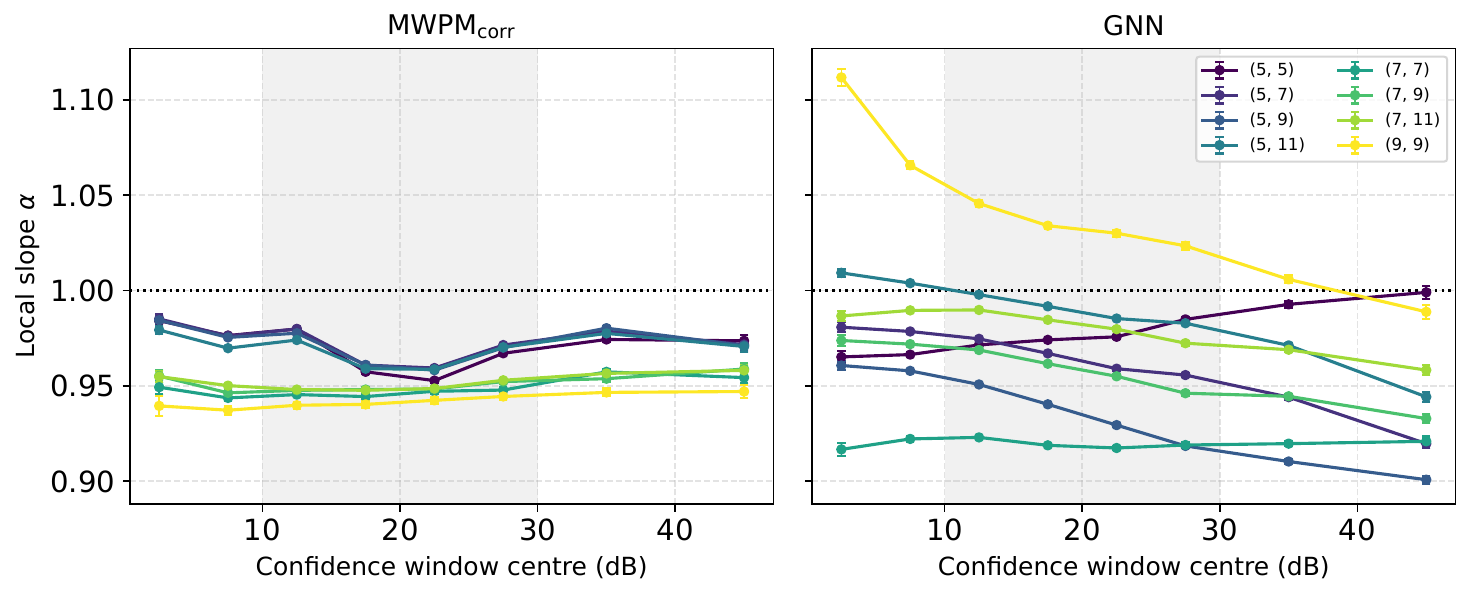}
    \caption{Local calibration slopes: Eq.~(\ref{true_gap_dB}) fitted in $5$-dB windows between $0$ and $30$ dB and in $10$-dB windows between $30$ and $50$ dB, plotted at the window centers, for MWPM$_{\rm corr}$ (left) and the GNN (right), for all eight configurations at $p = 5 \times 10^{-3}$. The shaded region is the baseline fit window, and the dotted line marks $\alpha = 1$. Error bars are unscaled Fisher standard errors (at most $0.0052$, largest in the outermost windows).}
\label{fig:local_slopes}
\end{figure*}

\emph{Local slopes.} Figure~\ref{fig:local_slopes} resolves the slope by confidence, fitting Eq.~(\ref{true_gap_dB}) in narrow windows across the range from $0$ to $50$ dB. For the correlated gap, the fitted local slopes vary only weakly with confidence at $d \geq 7$ ($0.937$--$0.959$ across all windows), consistent with an approximately constant rescaling over the examined range. At $d = 5$, the slopes range from $0.953$ to $0.985$ with a dip between $15$ and $25$ dB, inside the baseline fit window, consistent with the larger $\chi^2_\nu$ at this distance. For the GNN, the local slope drifts with the confidence in most configurations, mostly decreasing toward high confidence (from $0.96$ to $0.90$ at $(5, 9)$, and from $1.11$ to $0.99$ at $(9, 9)$), whereas it is nearly flat at $(7, 7)$ and increases at $(5, 5)$. The direction and magnitude of this drift differ between networks, while allowing an additional offset substantially improves the GNN calibration fits (\hyperref[app:calib]{Appendix~C}).

\begin{figure}
\centering
\includegraphics[width=0.7\textwidth]{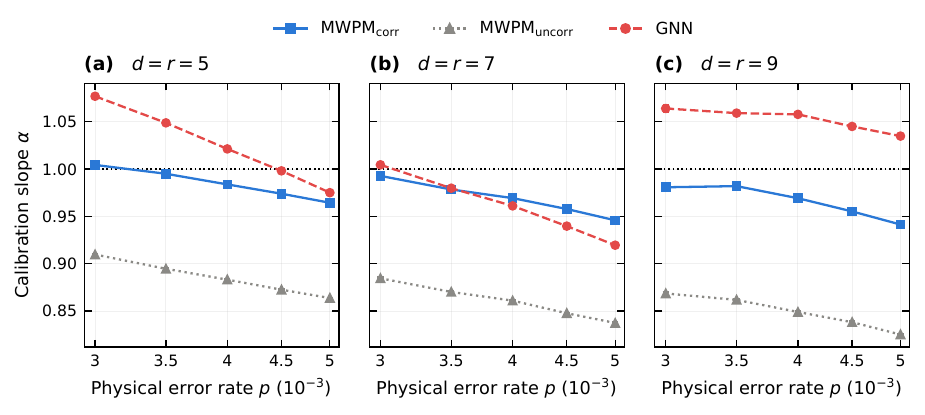}
    \caption{Calibration slope $\alpha$ as a function of the physical error rate $p$ for MWPM$_{\rm corr}$, MWPM$_{\rm uncorr}$ and the GNN, at (a) $d = r = 5$, (b) $d = r = 7$, and (c) $d = r = 9$, fitted as in Fig.~\ref{fig:gap2} (window $g \in [10, 30]$ dB), with $10^8$ shots per point. Unscaled Fisher uncertainties are smaller than the markers. The points at $p = 5 \times 10^{-3}$ are those of Table~\ref{alpha_table}, and the black dotted line marks $\alpha = 1$.}
\label{fig:alpha_vs_p}
\end{figure}

\emph{Physical error rate.} Fig.~\ref{fig:alpha_vs_p} shows the fitted slopes as a function of $p$ at $d = r = 5$, $7$ and $9$. For all three scores, $\alpha$ increases as $p$ decreases. For the GNN, the change over $p = 3$--$5 \times 10^{-3}$ is about $0.10$, $0.085$ and $0.03$ at $d = 5$, $7$ and $9$, and the slope crosses unity within the tested range at $d = 5$ and $7$; for both matching gaps it is about $0.04$--$0.05$ at every distance.
As discussed in the main text, the GNN is trained on a mixture of error rates, Eq.~(\ref{mixture}), without $p$ as an input, so that its calibration is expected to vary with the benchmark error rate, whereas the matching weights are always computed at the correct $p$. Throughout the tested range, the correlated gap has a calibration slope substantially closer to unity than the uncorrelated gap.

\section{Appendix F: Exact posterior at \texorpdfstring{$d = 3$}{d = 3}}
\phantomsection\label{app:exact}

For $(d, r) = (3, 3)$, the detector error model without decomposition has $219$ independent error mechanisms acting on $24$ detectors and the logical observable. Treating the observable as a $25$th bit, the joint distribution $P(s, L)$ over all $2^{25}$ syndrome-observable pairs follows from the Fourier (Walsh-Hadamard) representation of a product of independent parity flips: with $v_j \in \{0, 1\}^{25}$ the flip pattern of mechanism $j$ and $p_j$ its probability,
\begin{align}
    \hat{P}(k) = \prod_j (1 - 2 p_j)^{[k \cdot v_j \ {\rm odd}]}, \qquad P(x) = 2^{-25} \sum_k (-1)^{k \cdot x}\, \hat{P}(k),
\end{align}
which is evaluated with a fast Walsh-Hadamard transform. This gives the exact $\Lambda(s)$ at $p = 5 \times 10^{-3}$ and, repeating the transform at $p \in \{1, 2, 3, 4, 5\} \times 10^{-3}$ and restricting to non-trivial syndromes, the training-mixture posterior of Eq.~(\ref{mixture}). We sample $10^8$ shots at $p = 5 \times 10^{-3}$ with a fixed seed and keep the $57{,}444{,}877$ non-trivial ones, since trivial syndromes are not seen by the GNN. On these shots we compute $g_{\rm uncorr}$, $g_{\rm corr}$ and the logit of the $(3, 3)$ network, each signed toward its decision. For a score $\lambda$, the per-shot divergence is $D(s) = \sum_{L} P(L|s) \log_2 [P(L|s)/q(L|s)]$ with $q(L = 0|s) = \sigma(a \lambda(s))$; Table~\ref{exact_d3_table} lists its mean natively and for the global $a$ that minimizes it, together with the mean divergence from the mixture posterior.

\end{document}